\documentclass{article}

\usepackage[numbers]{natbib}
\usepackage{rotating}
\usepackage{graphicx}
\usepackage{url}

\begin{document}

\title{Supersymmetry: Experimental Status}

\author{Keith A. Ulmer, \\on behalf of\\
the CMS and ATLAS Collaborations \\ 
\\
Texas A\&M University\\ 
Department of Physics and Astronomy\\
College Station, TX, 77843, USA.}

\maketitle

\begin{abstract}
This talk presents results from the CMS and ATLAS Collaborations from searches for physics beyond the Standard Model motivated by supersymmetry from Run 1 of the
LHC. Representative searches are described to illustrate the diverse nature of the search program in both background estimation techniques and final state 
topologies. The status of preparation for Run 2 searches at 13 TeV is also presented.
\end{abstract}

\section{INTRODUCTION}
The Standard Model of particle physics accurately describes the interactions of all known fundamental particles in the universe, and has
remained the prevailing paradigm in the field for over forty years. Despite its success, the Standard Model remains an
incomplete theory of fundamental particles and interactions. It does not include a description of gravity, nor does it explain the compelling 
astronomical evidence for dark matter. Of the proposed extensions to the Standard Model, supersymmetry (SUSY) has remained among the most
popular for decades. It provides exactly the needed compensation to stabilize the Higgs mass, while additionally providing an ideal 
candidate for dark matter with a stable weakly interacting lightest supersymmetric particle (LSP). 

The CMS~\cite{CMS} and ATLAS~\cite{ATLAS} experiments at the CERN Large Hadron Collider are general purpose detectors built to explore 
the fundamental nature of the universe. Among the results from the two experiments are many searches for supersymmetry, which have thus
far yielded null results~\cite{CMS-SUSY, ATLAS-SUSY}. The search programs in both experiments are based on a wide arrange of techniques to 
measure standard model background contributions as well as a diverse range of possible final states. In this talk, a sample of results
are shown to illustrate techniques deployed in these searches. By no means are all relevant results discussed or presented.

In the first section, a series of general searches in different final states are described. The second section contains a discussion
of more targeted searches focused on dedicated final state topologies, while the third section discussed difficult to reach signatures.
The forth section attempts to put the full set of searches performed into a global context, while the final section shows progress
toward new searches in the LHC Run 2 with 13 TeV proton-proton collisions.

\section{INCLUSIVE SEARCHES}
Unlike the Standard Model Higgs boson, supersymmetry has many free parameters, which can give rise to a great variety of signatures. Further,
the unknown mass spectrum can also give rise to a great variety of production cross sections and final state kinematics. With such a broad
range of possible signatures, a fruitful class of supersymmetry searches is performed with inclusive sensitivity. Here I describe four such
examples in different final states.

The classic jets plus missing energy signature is searched for in a three dimensional binned analysis taking advantage of sensitivity in different
bins of missing energy (MET), the sum of jet transverse momenta (HT), and the number of jets tagged as bottom quarks~\cite{Chatrchyan:2013wxa}.
Events are selected by removing those with an identified electron or muon and then requiring at least three jets and at least one tagged
$b$-quark jet.
The main Standard Model backgrounds derive from $t\bar{t}$, $W$ + jets, $Z$ + jets, and QCD multijet events. The contribution from each category of 
background is measured from data control samples to minimize the reliance on accurate simulation. In particular, single lepton events are used
to predict the $t\bar{t}$ and $W$ + jets backgrounds, and dilepton events are used to predict the $Z$ + jets backgrounds with $Z$ decays to neutrinos.
The QCD multijet contribution is predicted by utilizing a kinematic sideband enriched in QCD events where a jet and MET are aligned.

\begin{figure}[!ht]
  \includegraphics[width=0.44\linewidth]{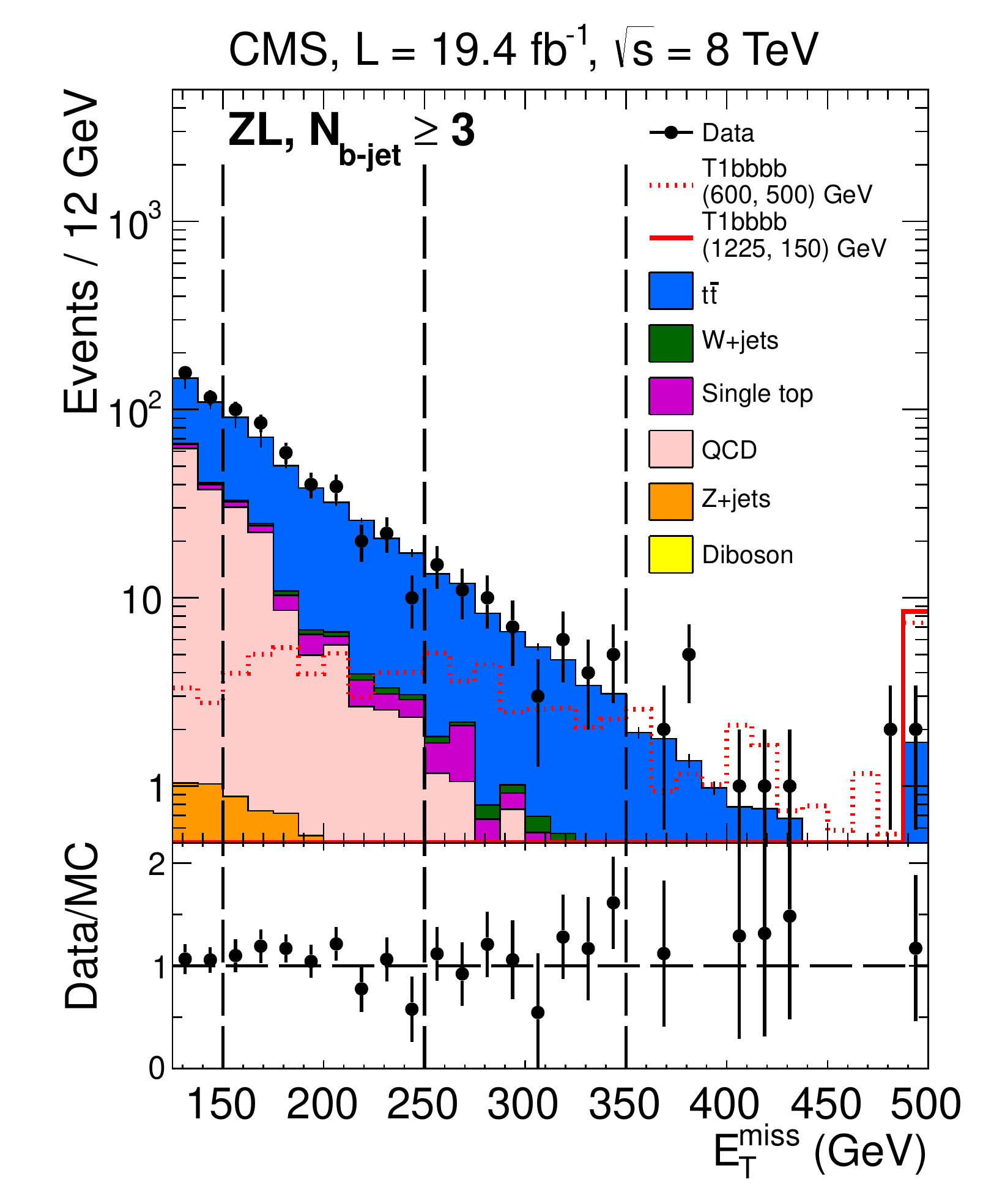}
  \includegraphics[width=0.54\linewidth]{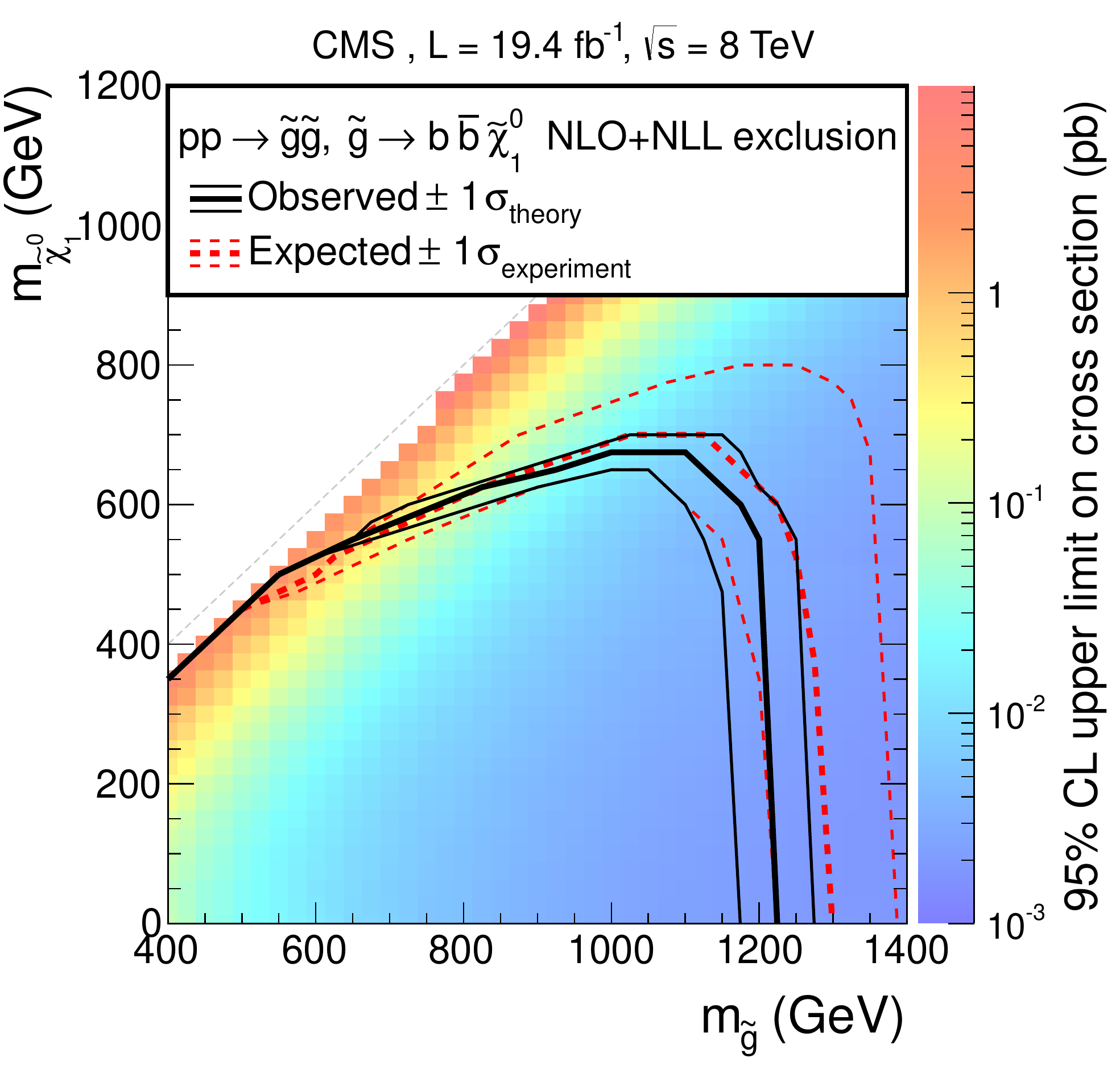}
  \caption{MET distribution in hadronic search with at least 3 $b$-jets (left) and limits from the search for gluino to $bb$LSP production (right).
  Taken from Ref.~\cite{Chatrchyan:2013wxa}.}
  \label{fig:hadronic}
\end{figure}

No significant excess of events above the Standard Model predictions is observed. Fig.~\ref{fig:hadronic} (left) shows the data compared to the
expected Standard Model contribution after selecting events with at least 3 $b$-quark jets. The search results are interpreted in several benchmark
SUSY models, including gluino pair production with each gluino decaying to two $b$-quarks and the LSP. Fig.~\ref{fig:hadronic} (right) shows that
such models are excluded for gluino masses as high as around 1.2 TeV. 

A similarly broad search was performed with complementary events selected with exactly one muon or electron in~\cite{Aad:2015iea}.
Minimal requirements on MET and the transverse mass (MT) of the lepton and MET are used to select SUSY-like events. 
Sensitivity to a variety of models is obtained by classifying search regions into large and small jet multiplicity. Backgrounds arise
predominately from $W$ + jets and $t\bar{t}$ events. The size of the Standard Model contributions are predicted by identifying data control
regions enriched in each background. MC simulation is then used with the overall normalization taken from the data control region to 
predict the background in each signal region. 

\begin{figure}[!ht]
  \includegraphics[width=0.42\linewidth]{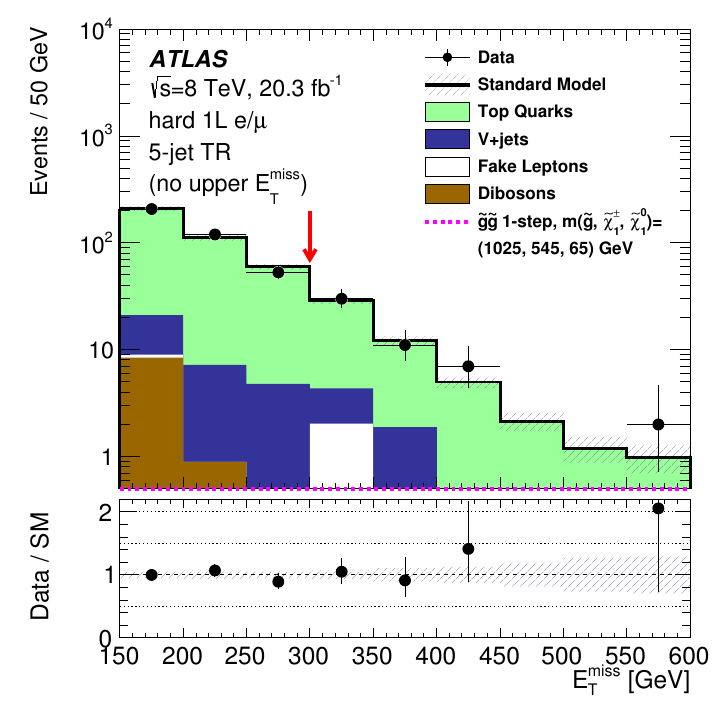}
  \includegraphics[width=0.57\linewidth]{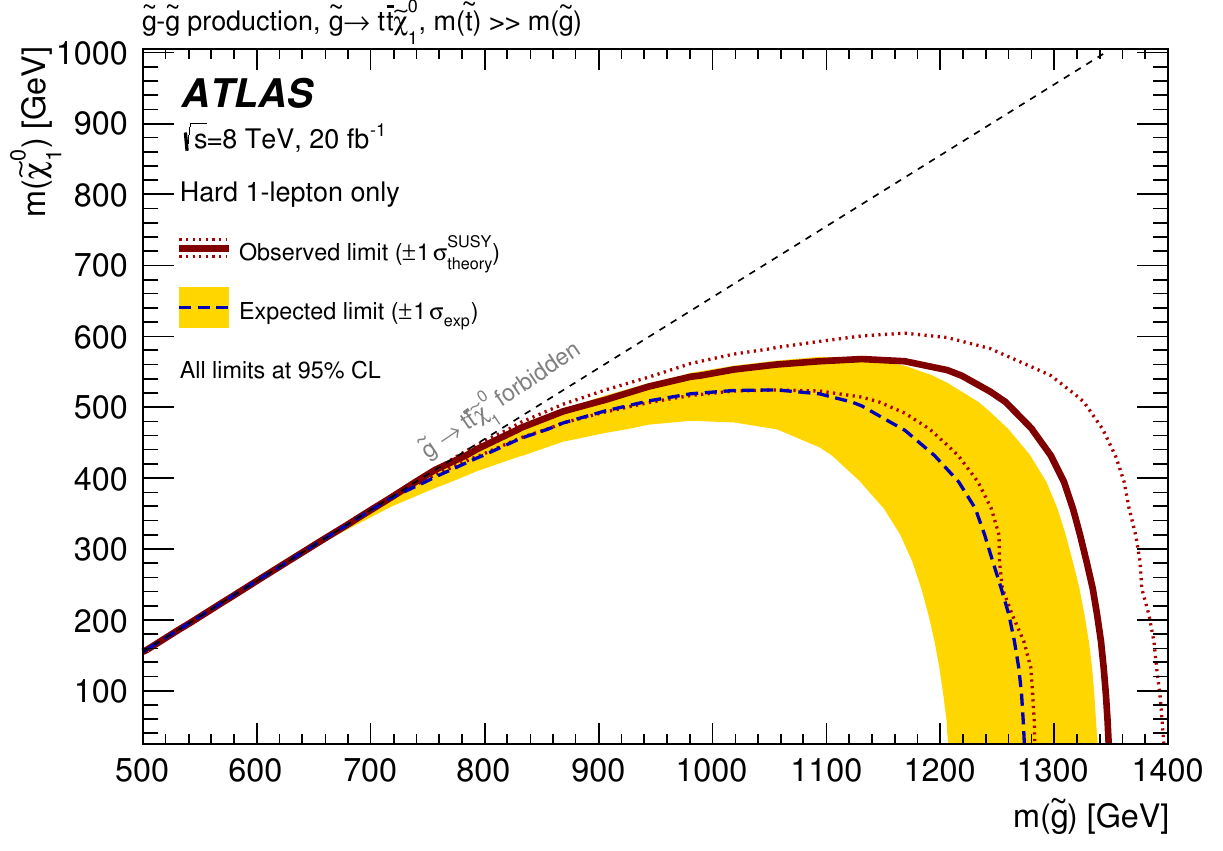}
  \caption{MET distribution for the $W$ + jets control region for the inclusive 1L search (left) and results from the search for gluino to $tt$LSP production (right).
  Taken from Ref.~\cite{Aad:2015iea}.}
  \label{fig:1L}
\end{figure}

The MET distribution for a signal region with five or more jets is show in Fig.~\ref{fig:1L} (left) with data compared to the background
estimation. No significant signal is observed in any of the search regions and 95\% CL upper limits are set on a variety of simplified 
models. Fig.~\ref{fig:1L} (right) shows the limits for gluino pair production with both gluinos decaying into two top quarks
with limits reaching beyond 1.3 TeV for light LSPs.

\begin{figure}[!ht]
  \includegraphics[width=0.52\linewidth]{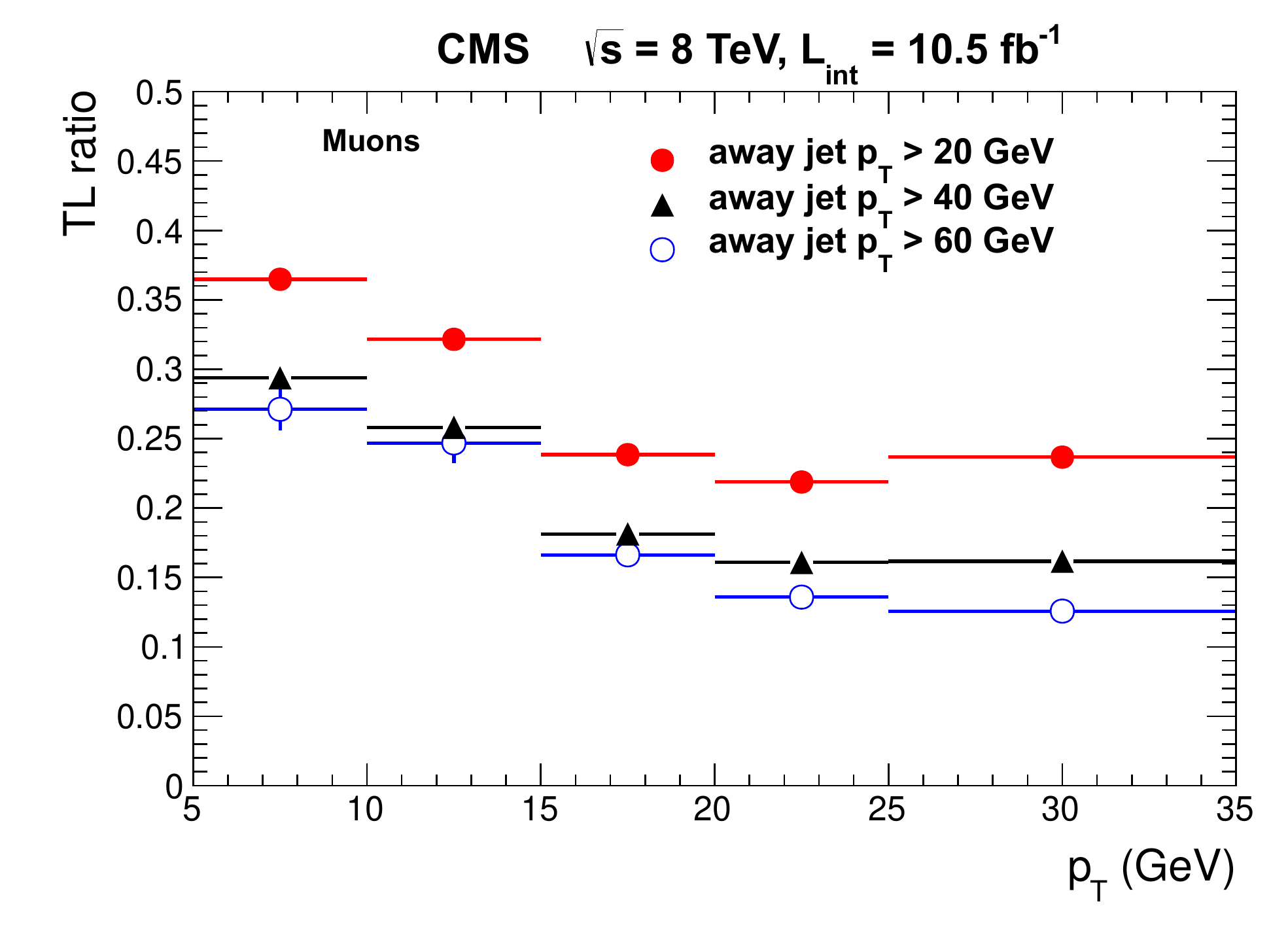}
  \includegraphics[width=0.46\linewidth]{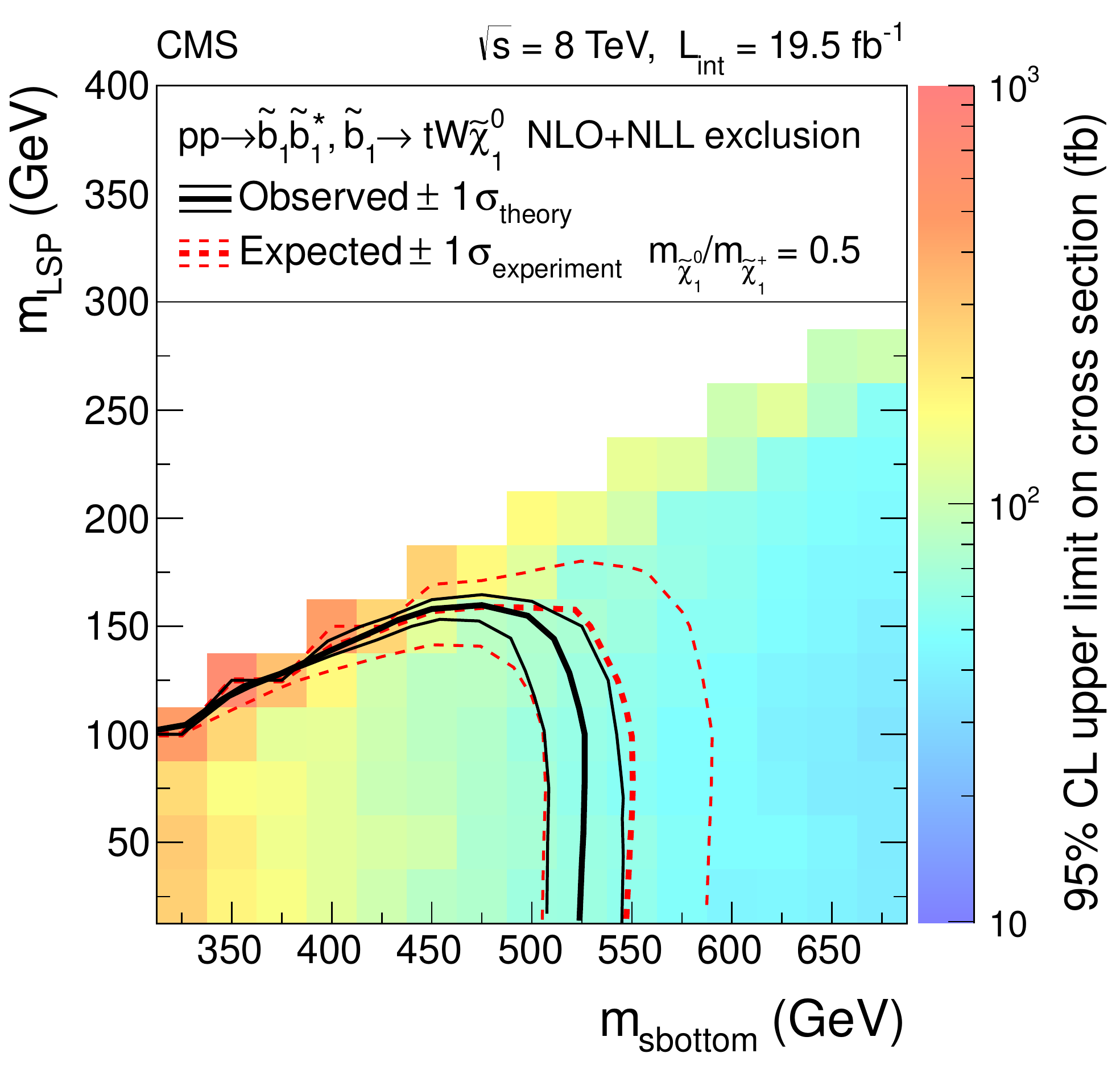}
  \caption{Tight to loose isolation ratio used to measure non-prompt lepton fake rate (left) and results from same-sign dilepton search for sbottom pair
  production (right). Taken from Ref.~\cite{Chatrchyan:2013fea}.}
  \label{fig:same-sign}
\end{figure}

Next, a search was performed based on events with two electrons or muons with the same electric charge~\cite{Chatrchyan:2013fea}.
Such events are rare in the Standard Model, but can occur readily in many new physics signatures. The leptons are required to be well isolated
to select prompt leptons from $W$ or $Z$ decays and remove those associated with jets, for example from semi-leptonic $b$ decays. 
The main backgrounds arise from events with a non-prompt lepton that mistakenly passes the isolation criteria in addition to another
prompt lepton or from events with two true prompt leptons arising from such rare processes as diboson production. The background from
non-prompt leptons is determined by measuring the so-called ``fake rate'' of the likelihood of a non-prompt lepton to pass the isolation
criteria in a data control sample, as shown in Fig.~\ref{fig:same-sign} (left). 
The contributions from rare backgrounds are taken from simulation. As with the other inclusive searches,
signal regions are defined in a number of bins of MET, HT, and number of $b$-tagged jets to ensure sensitivity to a variety of possible
signal models and parameter space. No significant excess of events is observed above the expected Standard Model backgrounds. 
Fig.~\ref{fig:same-sign} (right) shows the
observed limit for sbottom pair production with each sbottom decaying to $t$, $W$, and the LSP.

\begin{figure}[!ht]
  \includegraphics[width=0.44\linewidth]{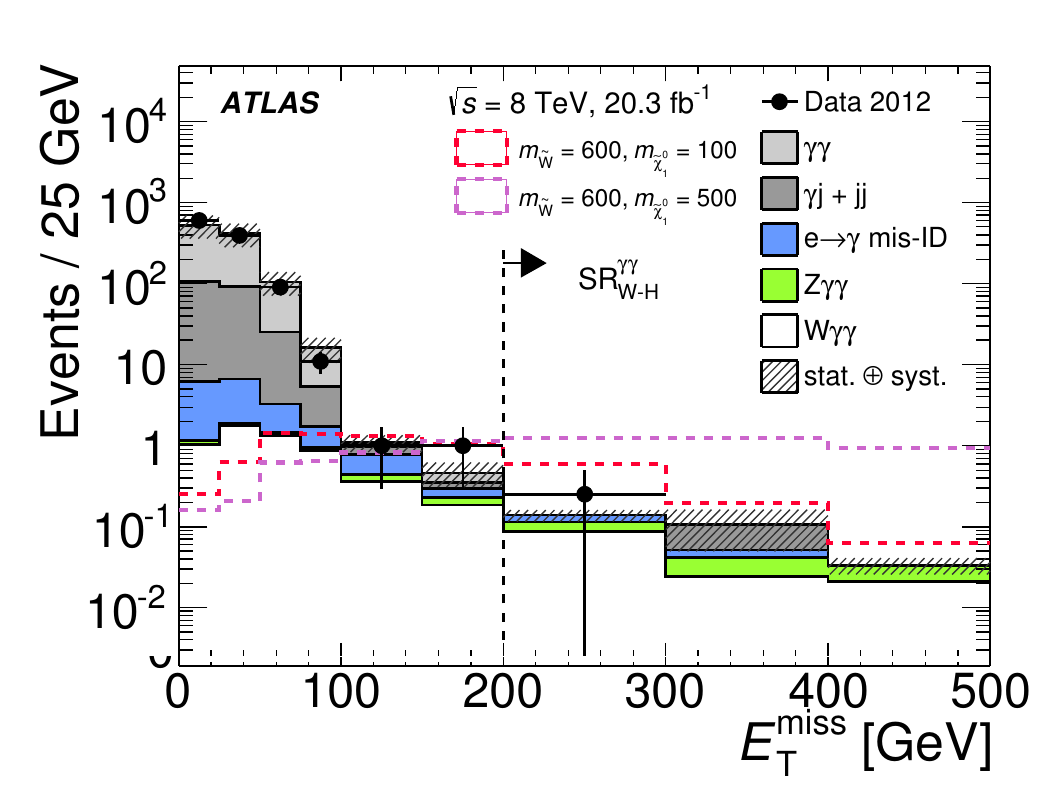}
  \includegraphics[width=0.52\linewidth]{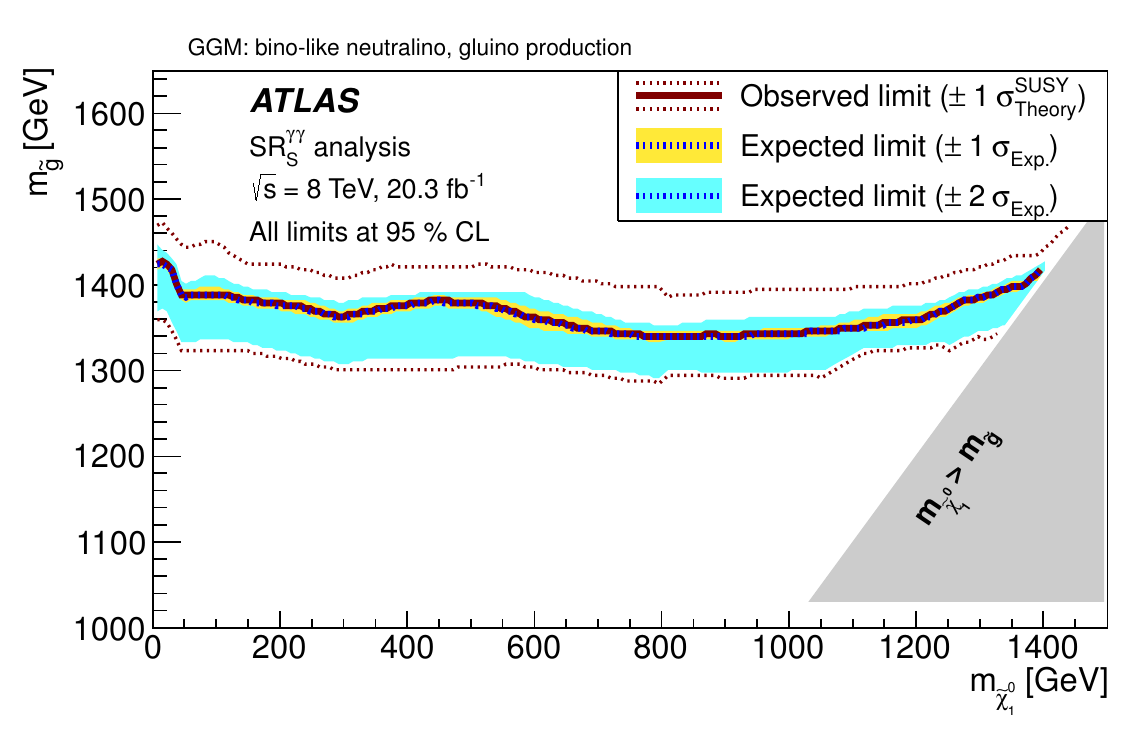}
  \caption{MET distribution for inclusive diphoton search signal region (left) and results for gluino production GMSB model (right).
  Taken from Ref.~\cite{Aad:2015hea}.}
  \label{fig:photon}
\end{figure}

The final inclusive search described in this talk is based on events with two high $p_T$ photons~\cite{Aad:2015hea}. 
Such a signature 
is common in gauge mediated (GMSB) SUSY scenarios where the LSP is a gravitino and the lightest neutralino decays
into a photon and the gravitino. Sensitivity to strong and electroweak production is achieved with search bins
in low and high MET and jet multiplicity. The main backgrounds arise from combinatorial diphoton production and
from photon + jets events where one of the jets is mistaken as an isolated photon. The missing energy in such 
background events is generally the result of mismeasured jets. Since the MET does not come from the photons 
themselves, data control samples composed of events from the photon isolation sidebands can be used to measure the
expected MET shape. The MET shape is then normalized to the low MET region in the true two photon sample to 
predict the background in the signal region. Fig.~\ref{fig:photon} (left) shows the MET distribution for a signal 
region with MET $>$ 200 GeV where the signal extends to higher MET than the remaining backgrounds. 
No significant excess of signal events is observed. Fig.~\ref{fig:photon} (right) shows the upper limits on 
gluino production in a GMSB scenario with a bino-like neutralino.
 
\section{TARGETED SEARCHES}
In addition to generic inclusive searches, some SUSY signatures are sufficiently well motivated to demand dedicated
searches targeting a more specific model. Here two such targeted searches are described. The first is
for stau pair production, while the second targets direct stop production.

\begin{figure}[!ht]
  \includegraphics[width=0.48\linewidth]{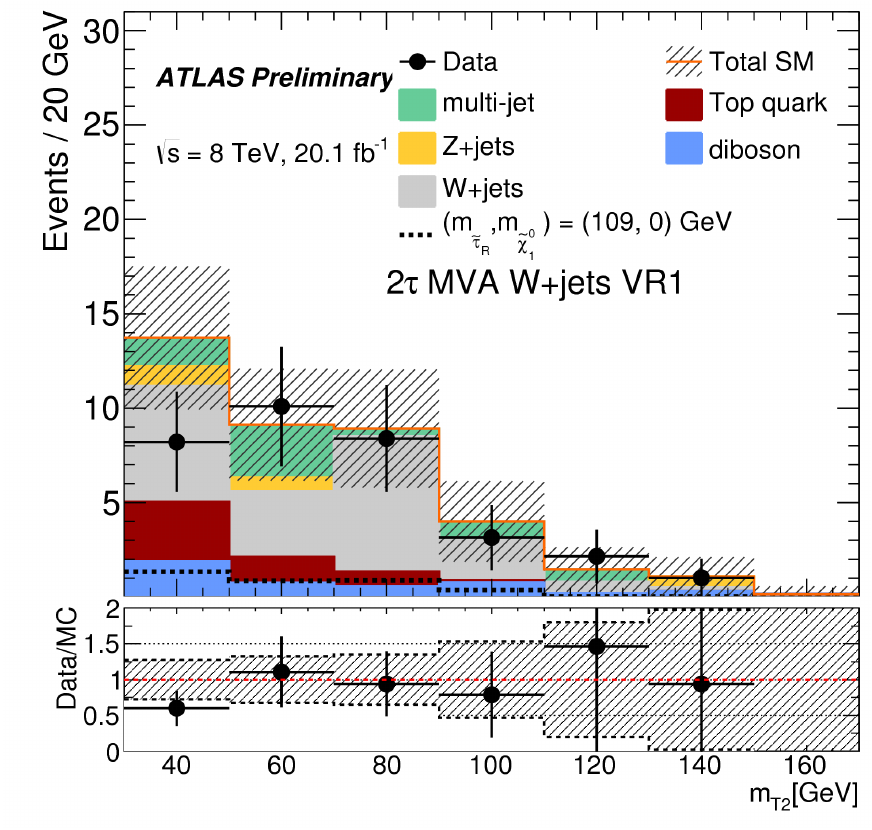}
  \includegraphics[width=0.48\linewidth]{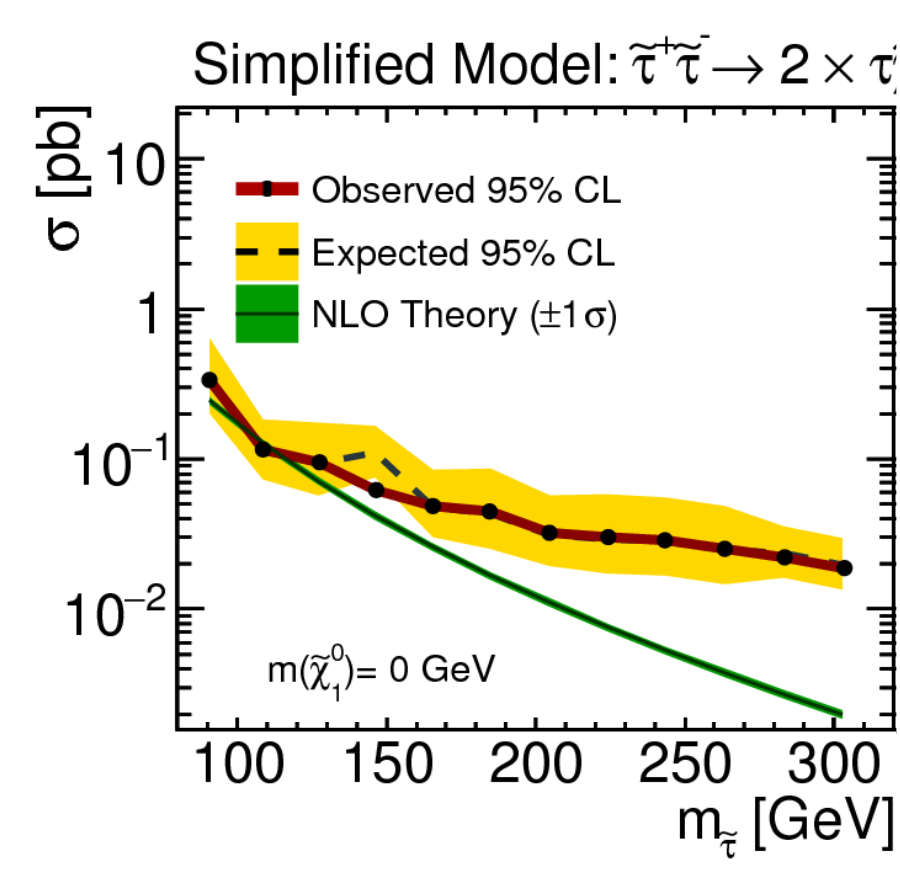}
  \caption{Stau search $W$ + jets validation region (left) and stau exclusion limits (right). Taken from Ref.~\cite{Aad:2015eda}.}
  \label{fig:stau}
\end{figure}

Direct stau pair production is well motivated, in particular by its potential connection to cosmological scenarios
to describe the early evolution of the universe. While generic dilepton searches are often sensitive to stau production
through the stau decays to electrons or muons, a dedicated search is required to capture sensitivity to hadronic stau 
decays which have the largest branching fraction. In~\cite{Aad:2015eda} events with two hadronic tau candidates with 
opposite charge are selected. $Z$ boson candidates are vetoed to reject $Z$ to $\tau\tau$ events, and events with
a $b$-tagged jet are rejected to remove $t\bar{t}$ events. The remaining background is dominated by QCD multijet events.
To select signal from this background, a multivariate boosted decision tree (BDT) is trained and only events with high 
BDT score are retained. After such selection, the main backgrounds remaining are $W$ + jets and diboson events. The
$W$ + jets background is measured by identifying a data control sample enriched in $W$ + jets and normalizing the MC
simulation prediction to the yield in this control sample. Fig.~\ref{fig:stau} (left) shows a plot of the MT2 distribution
in this control sample. No significant excess of events is observed about the background predictions and 95\% CL upper limits
are set. Even after stringent selection and significant background rejection, only the lightest staus are excluded with masses
around 100 GeV, as shown in Fig.~\ref{fig:stau} (right).

\begin{figure}[!ht]
  \includegraphics[width=0.44\linewidth]{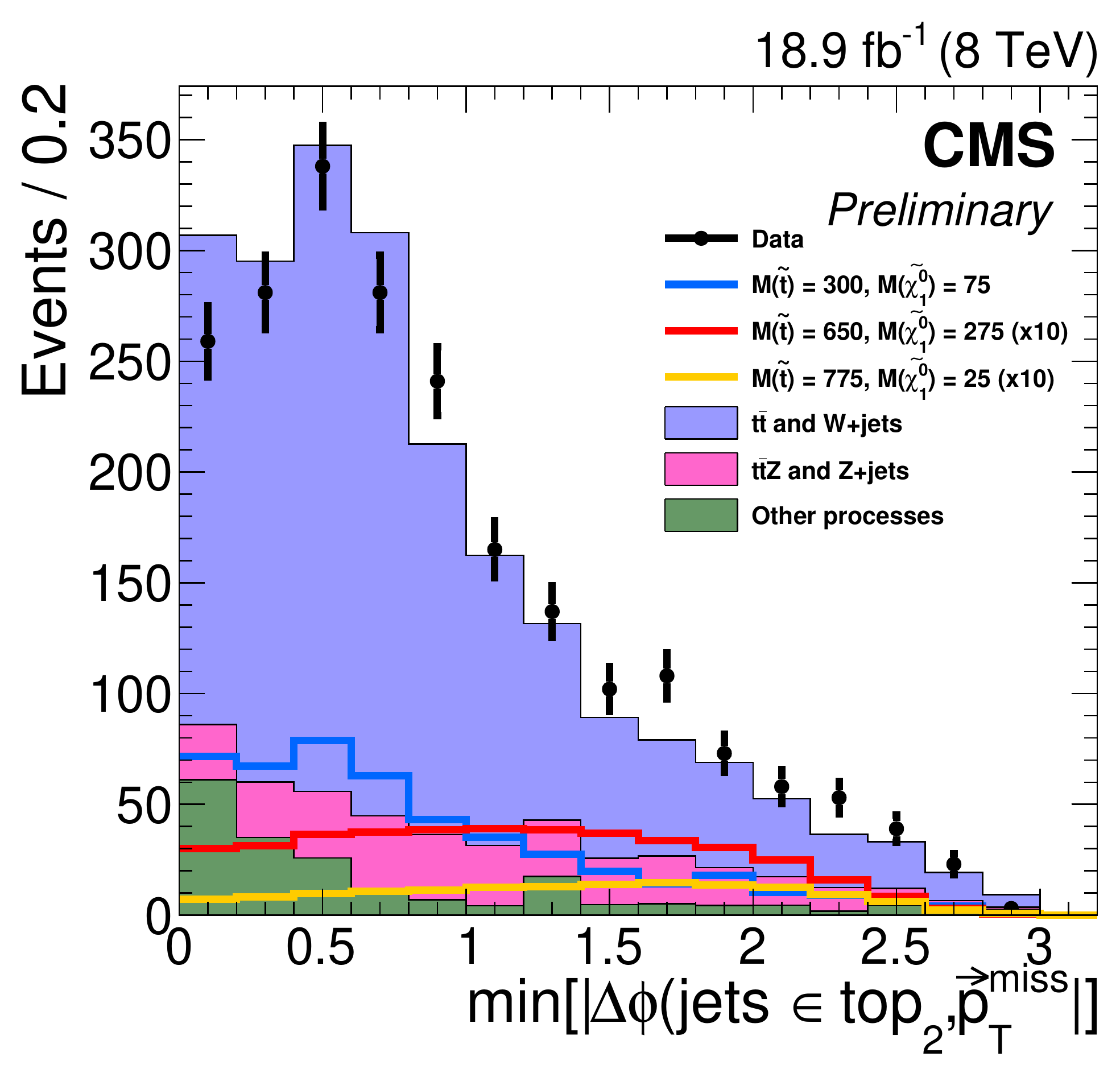}
  \includegraphics[width=0.52\linewidth]{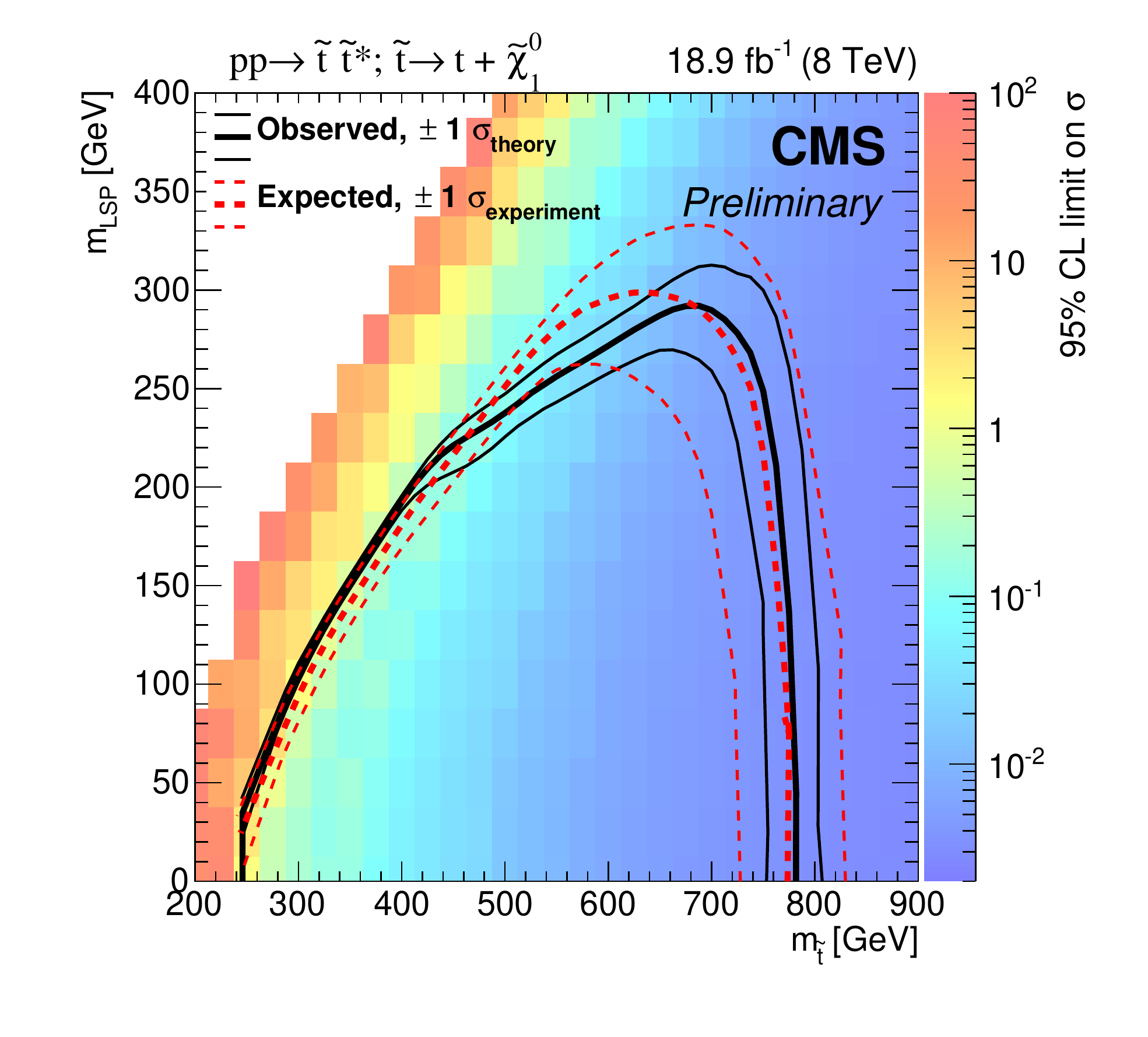}
  \caption{Minimum $\Delta\phi$ distribution between MET and subleading top candidate jets (left) and stop to top, LSP production limits (right)
  from the hadronic stop search. Taken from Ref.~\cite{CMS-stop}.}
  \label{fig:stop}
\end{figure}

Another very well motivated SUSY scenario is direct stop production, as the stop plays a key role in the cancelation of quadratic 
divergences to the Higgs mass from top quark loops. The search in~\cite{CMS-stop} utilizes the all hadronic final state to target
stop pair production with stop to top, LSP decays with both tops decaying hadronically. The analysis uses a customized jet
algorithm to identify two hadronic top decay candidates. Events with an isolated electron, muon, or tau are removed. The most
significant background arrises from $t\bar{t}$ events with MET from a leptonic $W$ decay where the charged lepton is lost.
The separation of signal and background is achieved with a BDT trained to select signal events. Kinematic variables such
as the angle between the MET and the jets in the sub-leading top candidate as shown in Fig.~\ref{fig:stop} (left) are used in the 
BDT. After the selection, MC simulation is used to estimate the total background contribution in each signal region. The MC is 
corrected to achieve good agreement with data in several key kinematic distributions and the background prediction is validated
in the BDT sidebands. No significant excess of data over background is observed and limits are set on direct stop production.
As shown in Fig.~\ref{fig:stop} (right) stop masses up to around 800 GeV are excluded for light LSPs.

\section{EXPLORING GAPS}
As more and more SUSY searches have yielded null results, an increasing effort has been placed on considering where a signal
may yet be hiding in space accessible with current LHC data. This section describes four such searches that explore regions
not covered by more conventional SUSY searches. 

\begin{figure}[!ht]
  \includegraphics[width=0.48\linewidth]{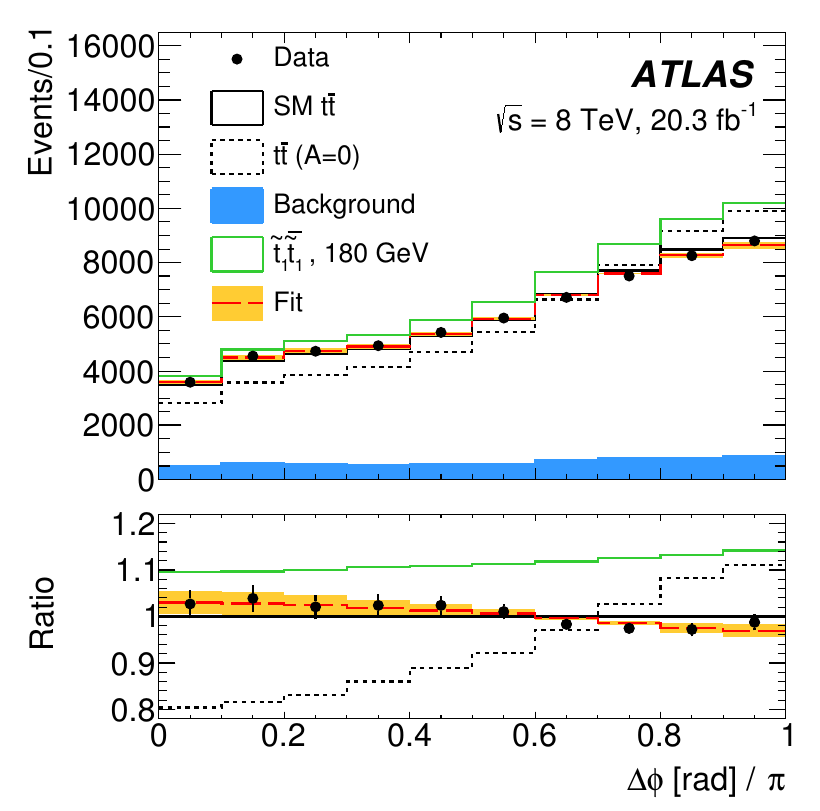}
  \includegraphics[width=0.48\linewidth]{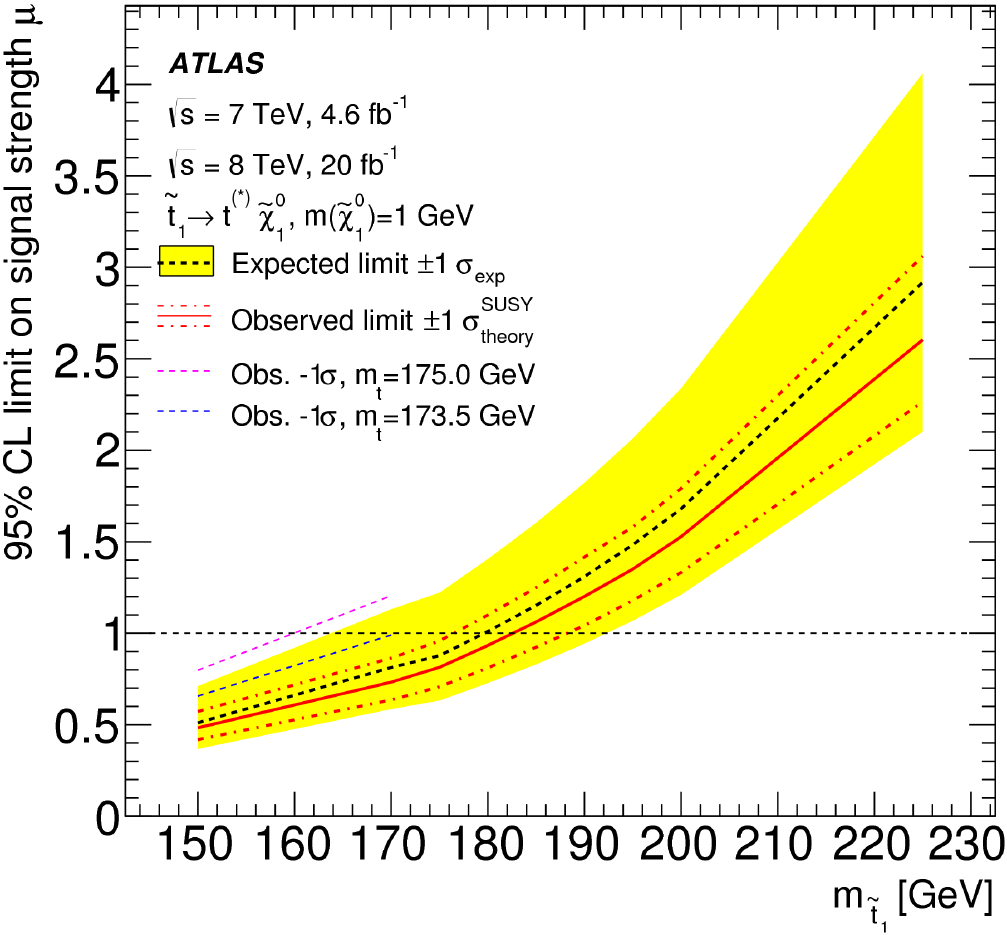}
  \caption{Angle between two leptons in dileptonic $t\bar{t}$ cross section measurement compared to SUSY stop signal (left)
  and resulting exclusion limit for stop production (right). Taken from Ref.~\cite{Aad:2014mfk}.}
  \label{fig:top}
\end{figure}

One such gap in SUSY sensitivity occurs when the stop has a mass very close to that of the top and LSP is very light.
In this scenario, stop pair production looks very similar kinematically to top pair production and the signal can be
very difficult to dig out. One approach is to use a precision measurement of the top cross section and compare it to
the theoretically predicted cross section from the Standard Model. If excess events exist, they could be from the presence 
of stops. Additionally, the spin correlations of the scalar stops are somewhat different from that of the spin 1/2 tops. 
The search in~\cite{Aad:2014mfk} exploits this difference to gain sensitivity to stop production in this difficult region.
Dileptonic $t\bar{t}$ events are used to compare the observed angular difference between the leptons with that expected from
$t\bar{t}$ and stop pair production, as shown in Fig.~\ref{fig:top} (left). No deviation from the expected Standard Model
distribution is observed and limits are set on stop pair production, as shown in Fig.~\ref{fig:top} (right).

\begin{figure}[!ht]
  \includegraphics[width=0.44\linewidth]{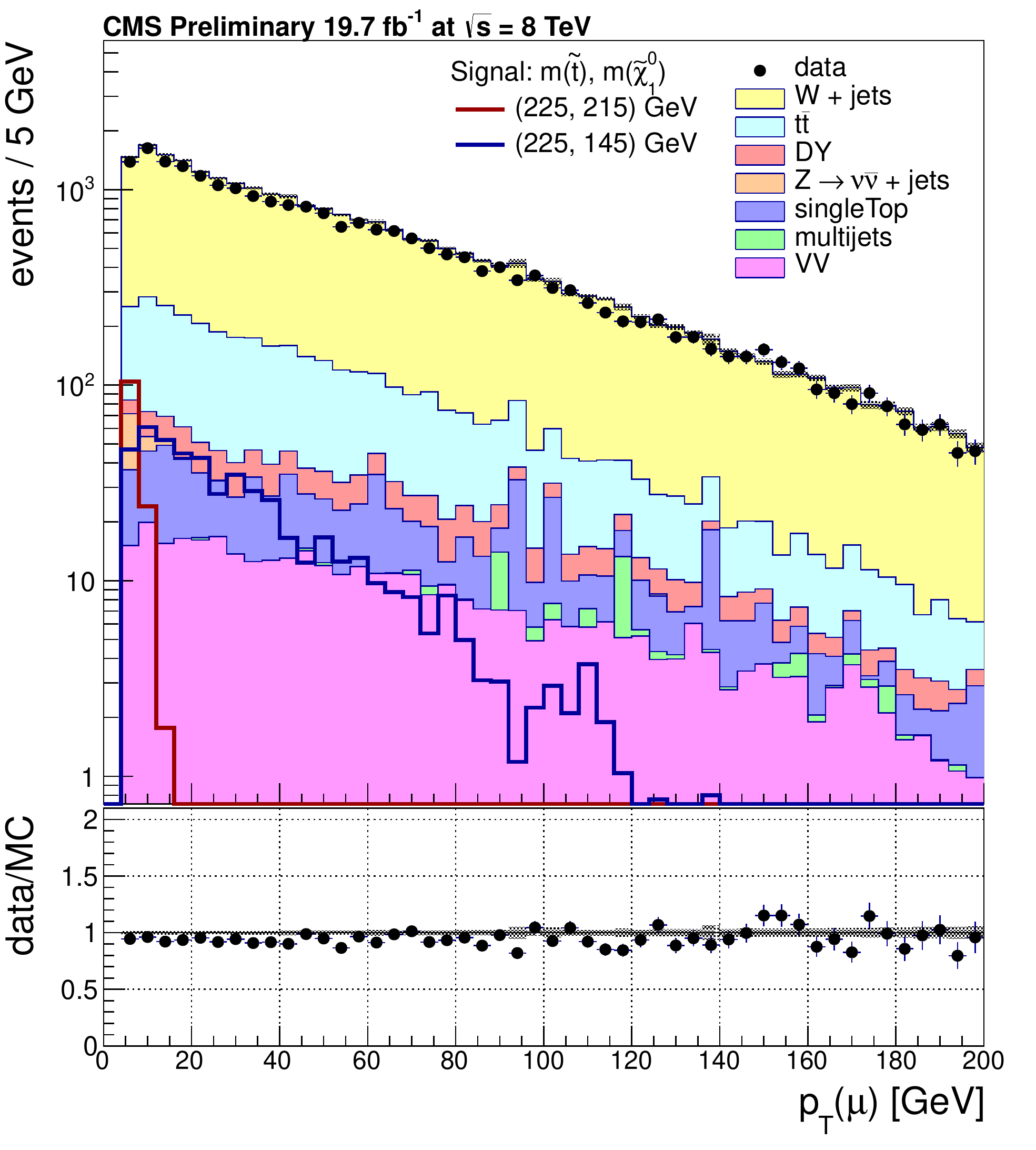}
  \includegraphics[width=0.52\linewidth]{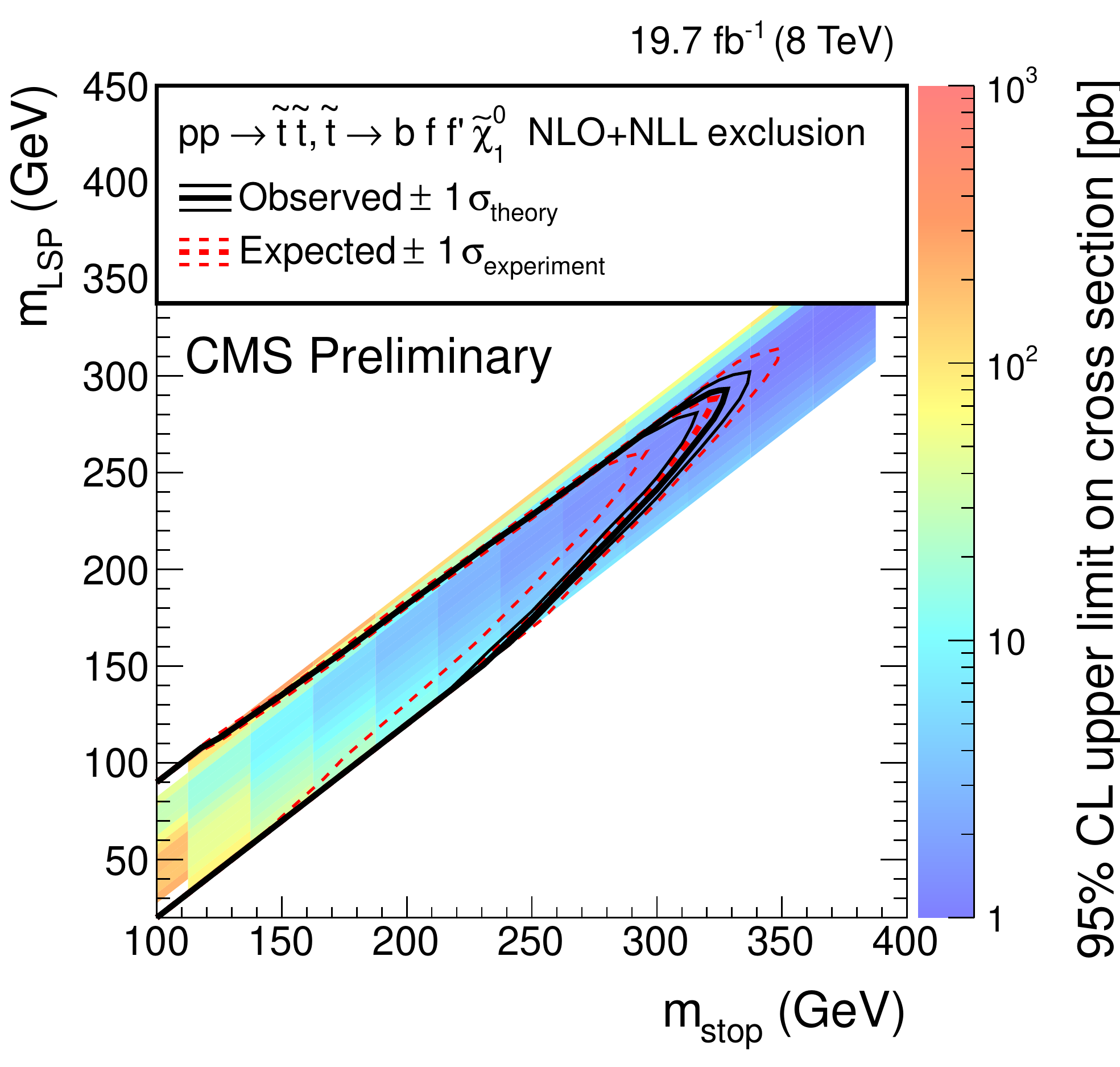}
  \caption{Muon $p_T$ distribution (left) and compressed stop search limits (right) in soft muon SUSY search. Taken from Ref.~\cite{CMS-softL}.}
  \label{fig:soft1L}
\end{figure}

Another difficult to access region occurs when SUSY particle masses are nearly degenerate. These so called ``compressed'' spectra
can result in SUSY decays with little missing energy if the LSP is close in mass the parent particle. As in direct dark matter 
searches with the monojet topology, compressed SUSY can be searched for in events where the SUSY system recoils against an ISR jet.
The compressed spectrum then produces missing energy when it is boosted. Such a technique is employed in~\cite{CMS-softL} where the
ISR jet and missing energy are searched for in combination with one or two low $p_T$ leptons, which can originate from stop or
chargino decays. The resulting lepton $p_T$ spectrum is soft, as shown in Fig.~\ref{fig:soft1L} (left) for compressed decays. After
selecting only events with low $p_T$ isolated leptons much of the background is removed and sensitivity to this difficult region 
is obtained, as shown in Fig.~\ref{fig:soft1L} (right).

\begin{figure}[!ht]
  \includegraphics[width=0.48\linewidth]{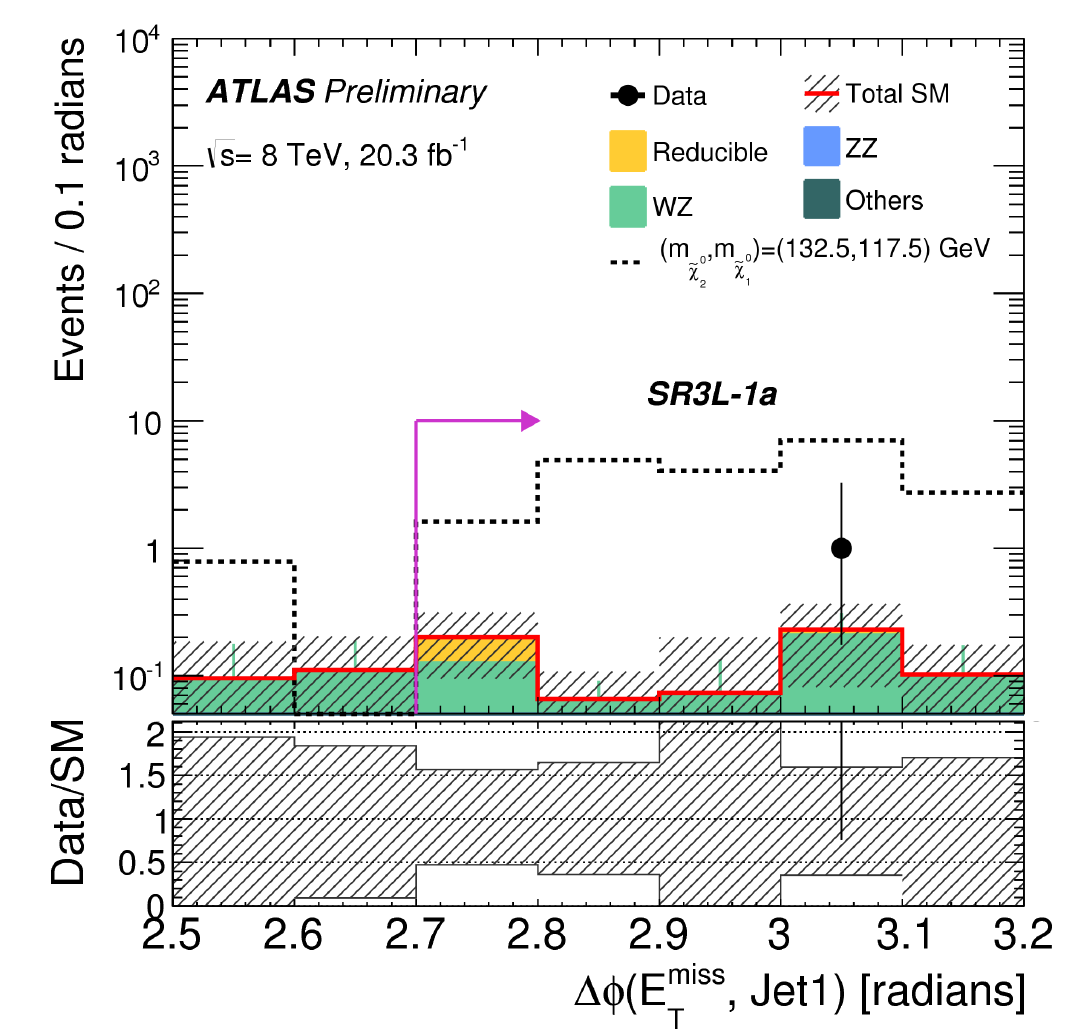}
  \includegraphics[width=0.48\linewidth]{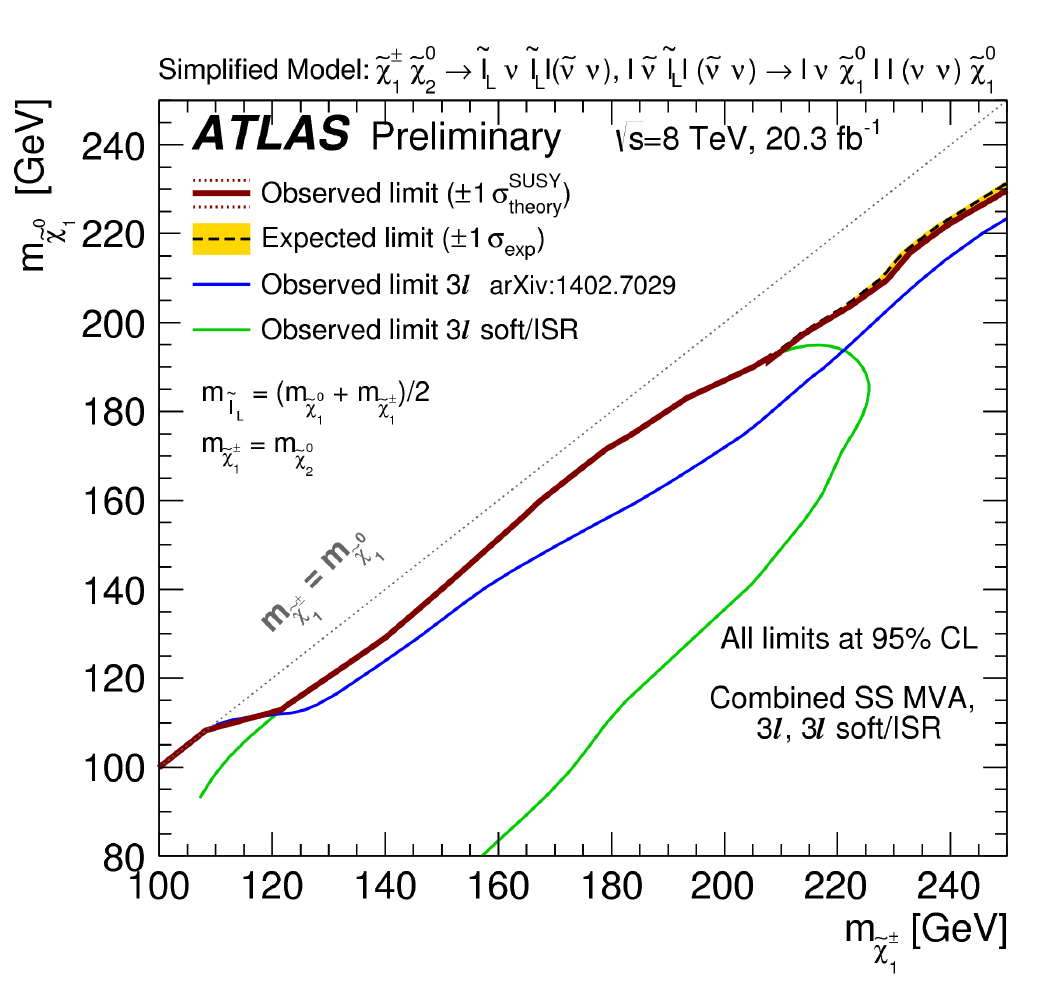}
  \caption{Angular separation between lead jet and MET in the three soft lepton + ISR search (left) and resulting
  limits on electroweak SUSY production (right). Taken from Ref.~\cite{Aad:2015eda}.}
  \label{fig:multiLep}
\end{figure}

The search in~\cite{Aad:2015eda} extends the soft lepton plus ISR topology even further in searching in events
with three or more low $p_T$ leptons plus large MET. This allows for sensitivity to such SUSY signatures as
chargino or neutralino production decaying to a neutralino LSP with intermediate sleptons, which can give up 
to four leptons in the final state. The Standard Model background for three or more isolated leptons plus large
MET and a high $p_T$ ISR jet is very low. Figure~\ref{fig:multiLep} (left) shows the single observed signal event
in one of the search regions compared to the background prediction, while Fig.~\ref{fig:multiLep} (right) shows 
the results of the search when combined with same-sign dilepton and high $p_T$ multilepton searches.

\begin{figure}[!ht]
  \includegraphics[width=0.38\linewidth]{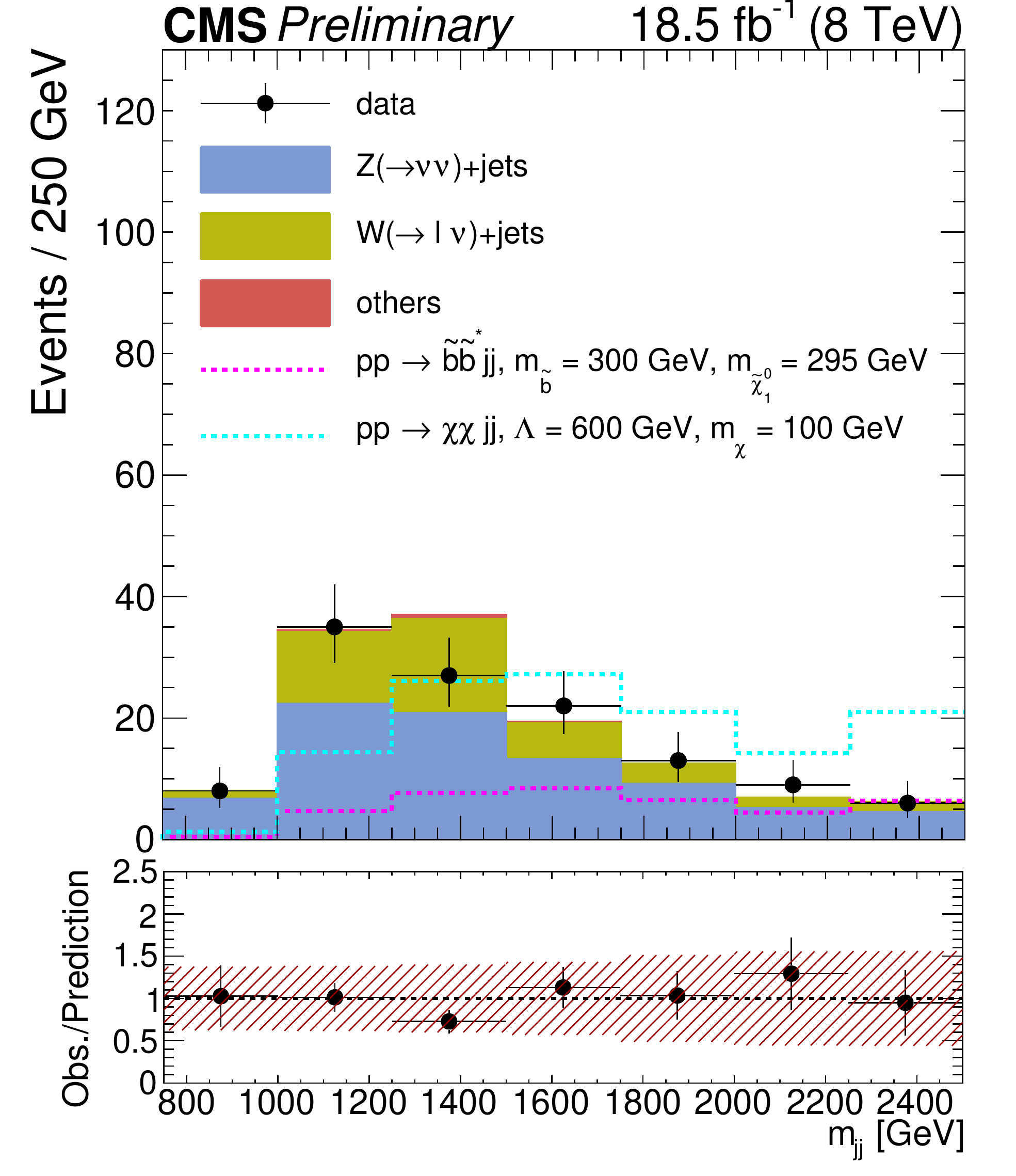}
  \includegraphics[width=0.58\linewidth]{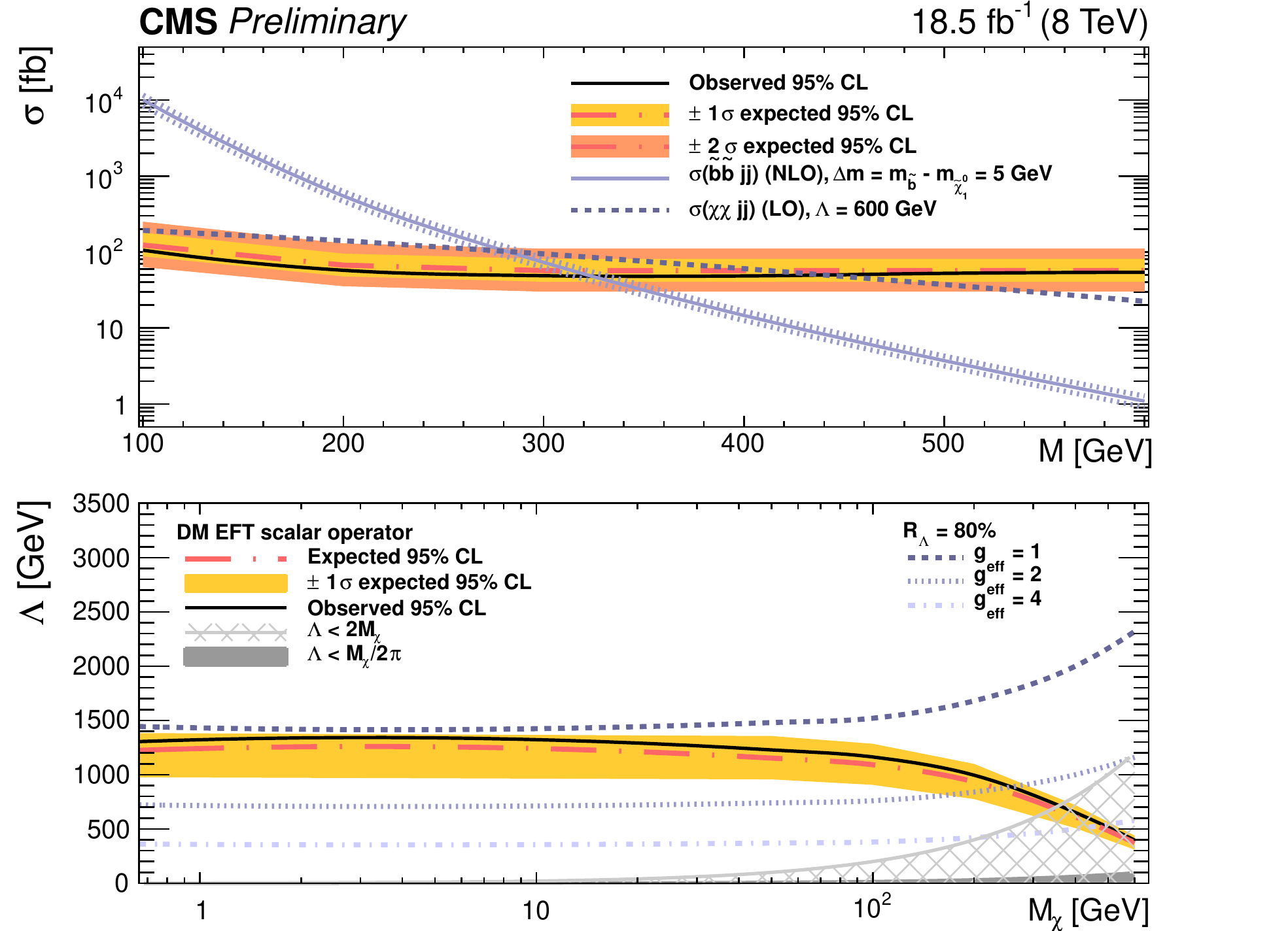}
  \caption{Dijet mass distribution comparing signal and background (left) and limits for sbottom pair
  and dark matter production (right) from the VBF SUSY search. Taken from Ref.~\cite{CMS-VBF}.}
  \label{fig:VBF}
\end{figure}

Another alternative to ISR to boost the compressed SUSY spectrum is vector boson fusion (VBF). The pair of VBF
jets serves the same purpose of providing a boost to the SUSY system, which would otherwise have very low MET.
The search in~\cite{CMS-VBF} exploits the VBF topology to search for compressed
SUSY with complementary sensitivity to the ISR searches. As an additional discriminating variable, the mass of the
VBF dijet system can be utilized to select high mass events more typical of signal. Figure~\ref{fig:VBF} (left) shows
the dijet mass distribution for background compared to signal. The observed distribution is consistent with the
Standard Model expectation and no evidence for SUSY is found. Figure~\ref{fig:VBF} (right) shows the search results
interpreted as limits on compressed sbottom pair production as well as direct dark matter production.

\section{THE BROAD PICTURE}

\begin{figure}[!ht]
  \includegraphics[width=0.48\linewidth]{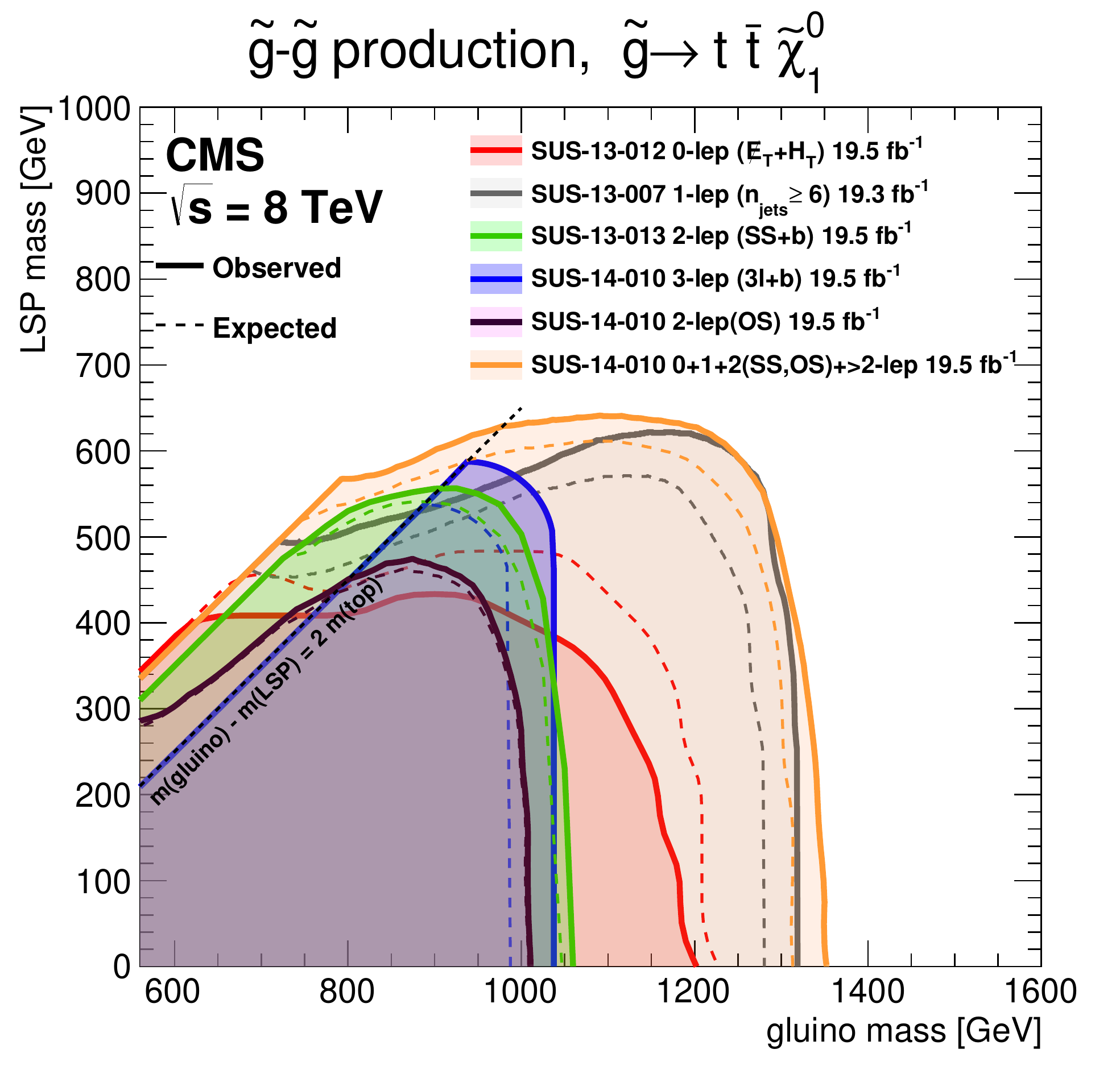}
  \includegraphics[width=0.48\linewidth]{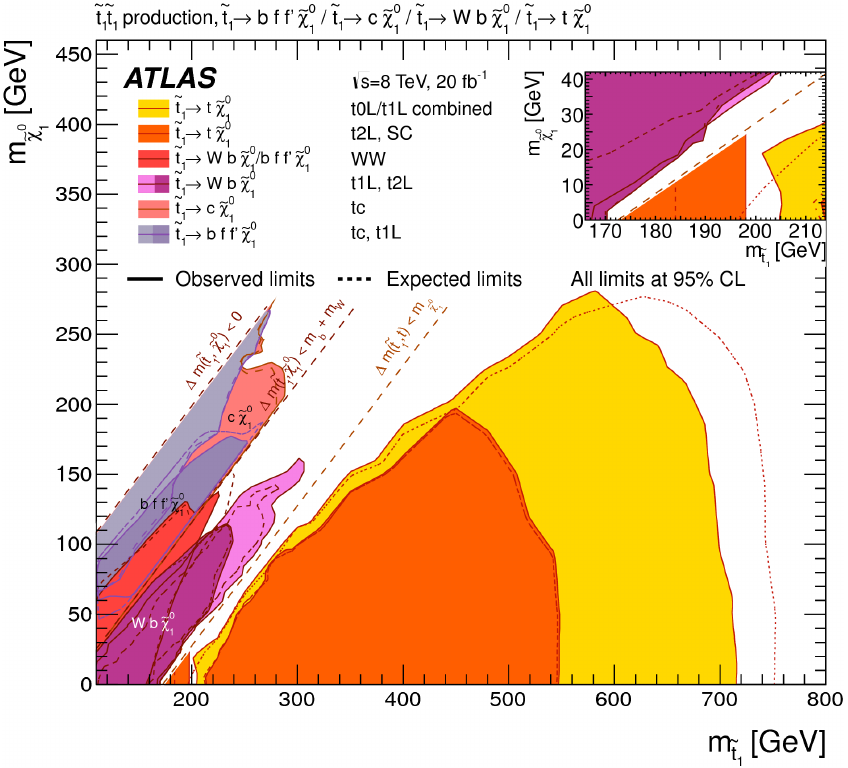}
  \caption{Summary plots of various search results from gluino-mediated stop production (left) from~\cite{CMS-SUSY} and direct stop production (right)
  from~\cite{ATLAS-SUSY}.}
  \label{fig:summaries}
\end{figure}

With the plethora of possible SUSY signatures and searches performed at the LHC, it is important to put the entirety of the search program
together to assess where things stand. Many different searches can be sensitive to the same model. When mutually exclusive final states
provide complementary sensitivity, a combination of the results of the different relevant searches can extend the overall reach. For example,
in Fig.~\ref{fig:summaries} (left) the results from searches using five different final states are shown along with the combination of the five
searches, which extends the sensitivity beyond any of the individual searches alone. Alternatively, different searches can be designed to be
sensitive to different regions of parameter space for a given model. When the exclusion regions for each individual search are overlaid, the total
exclusion can show significant coverage. For example, Fig.~\ref{fig:summaries} (right) shows the exclusions from eight different searches
targeting direct stop production. In total, they exclude a very significant region of the plane. Such summary plots also serve to highlight 
regions where gaps exist in the current sensitivity and can motivate future efforts.

\begin{figure}[!ht]
  \includegraphics[width=0.48\linewidth]{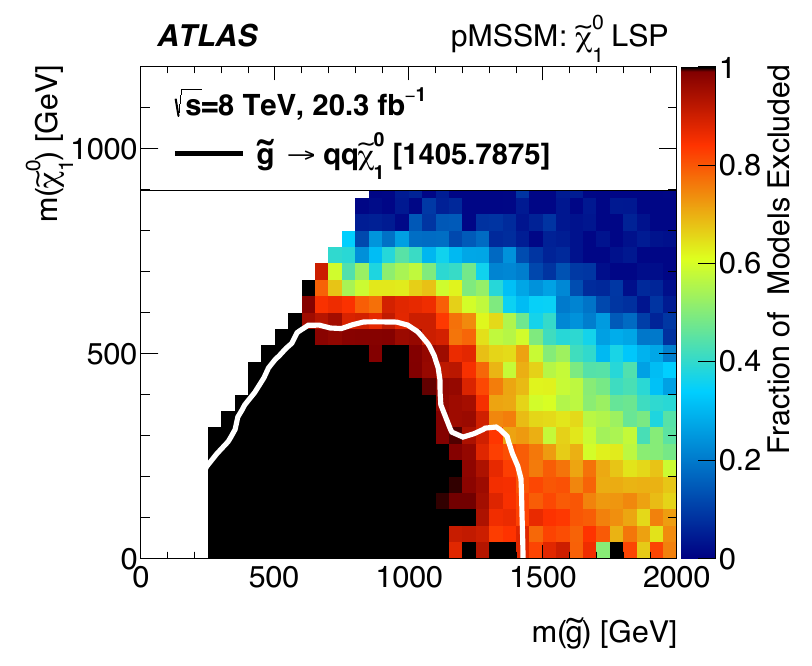}
  \includegraphics[width=0.48\linewidth]{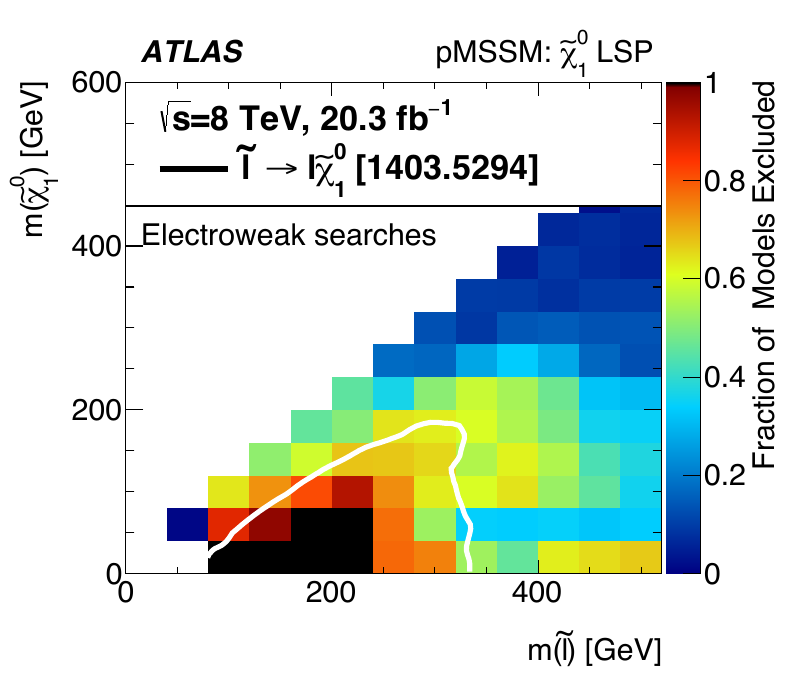}
  \caption{Results for gluino (left) and slepton (right) exclusions from the pMSSM parameter scan points. Taken from Ref.~\cite{Aad:2015baa}.}
  \label{fig:pMSSM}
\end{figure}

An alternative approach to assess the overall state of the SUSY search program is to consider full SUSY models. A popular approach is
to utilize the parameterized minimal supersymmetric standard model (pMSSM) which parameterizes SUSY with 19 free parameters after making
several experimentally well motivated assumptions. Many SUSY signal points are then generated based on scanning the 19 parameters to provide
a set of possible SUSY mass spectra. In~\cite{CMS-pMSSM} and~\cite{Aad:2015baa} scans of pMSSM points are compared to a variety of results
from the CMS and ATLAS collaborations, respectively. The points are then classified into those which are excluded by at least one of the 
searches and those that remain viable. The fraction of excluded points for gluino and slepton production from~\cite{Aad:2015baa}, for 
example, are show in Fig.~\ref{fig:pMSSM}. As expected, the lower mass points are more likely to be excluded and the results generally compare
well to the simplified model results. However, certain of the pMSSM points that remain allowed can be studied in further detail to understand
how to better design searches to capture sensitivity to these points in future searches.

\section{PREPARATION FOR 13 TeV}
The year 2015 saw the restart of the LHC after ``long shutdown 1'' in 2013-2014. The shutdown allowed for the successful
retuning of the LHC to achieve a record collision energy of $\sqrt{s} = 13$ TeV. At the time of the LHCP conference,
each experiment had collected a few dozen pb$^{-1}$ worth of 13 TeV data, which was used to commission the 13 TeV 
SUSY searches~\cite{SUS-Commissioning-DPS,ATLAS-Z,ATLAS-QCD}. In this section, results of these commissioning exercises are shown. 

\begin{figure}[!ht]
  \includegraphics[width=0.48\linewidth]{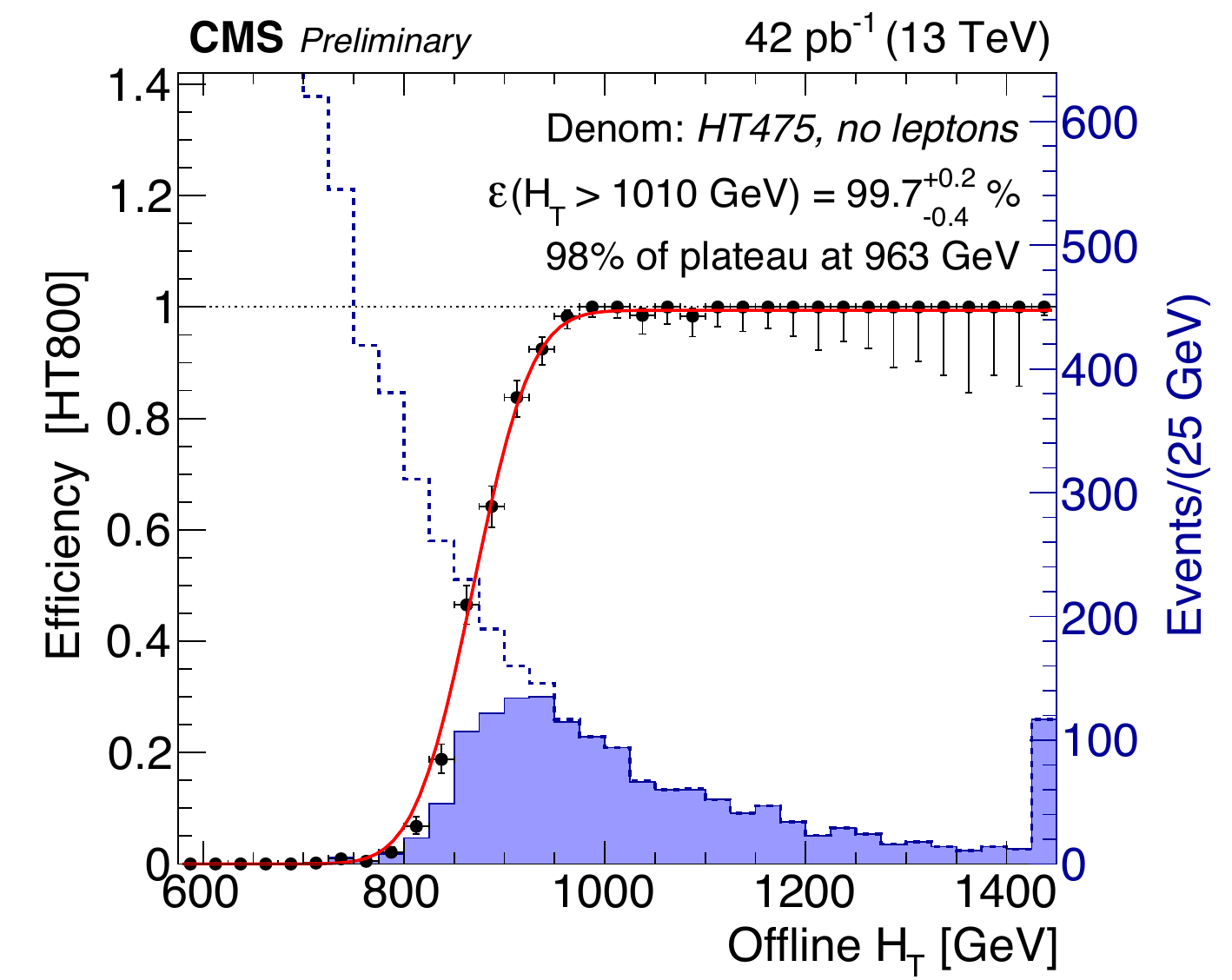}
  \includegraphics[width=0.48\linewidth]{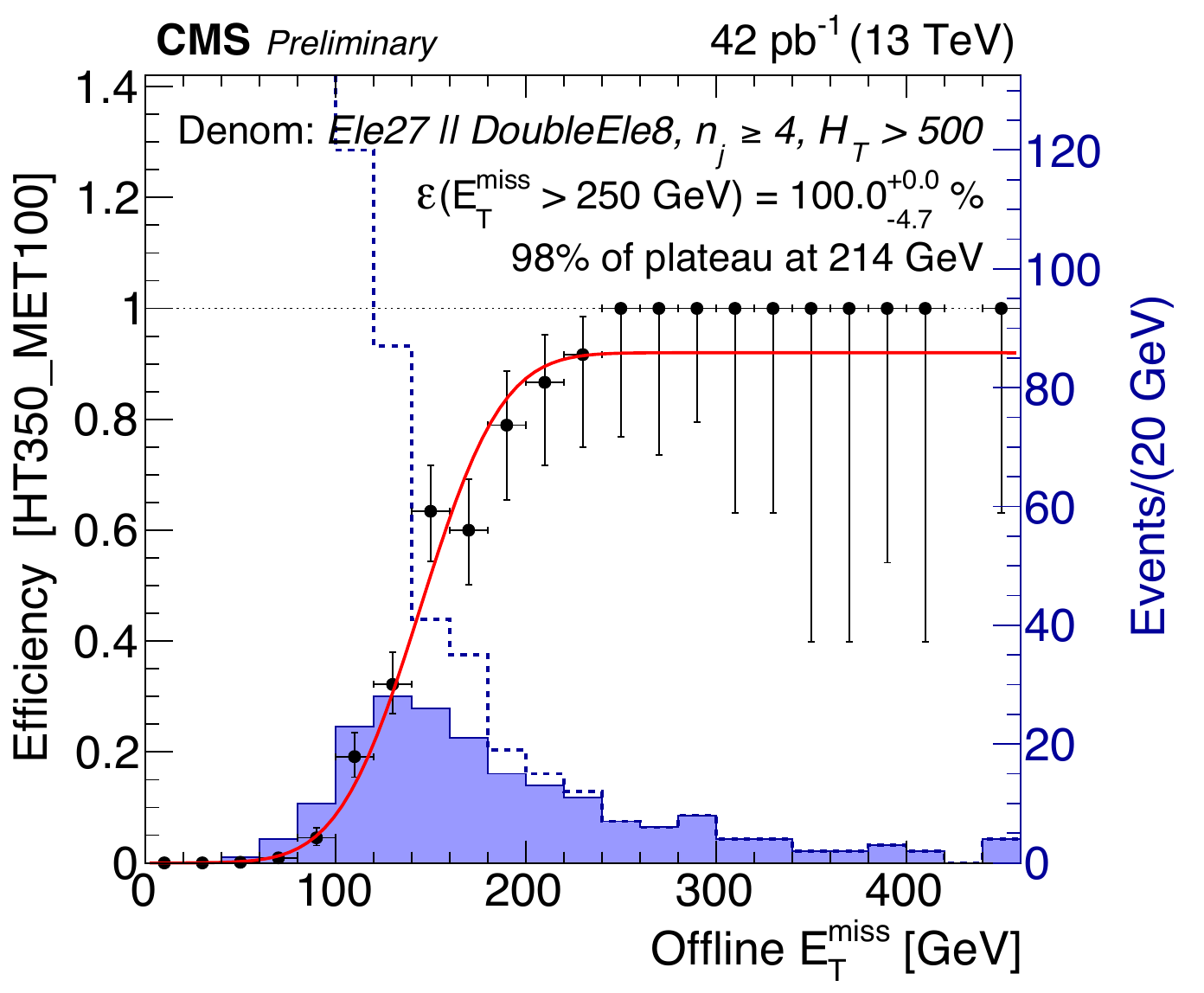}
  \caption{Early 13 TeV data commissioning plots for SUSY triggers based on HT (left) and MET (right). 
  Taken from Ref.~\cite{SUS-Commissioning-DPS}.}
  \label{fig:13trig}
\end{figure}

\begin{figure}[!ht]
  \includegraphics[width=0.48\linewidth]{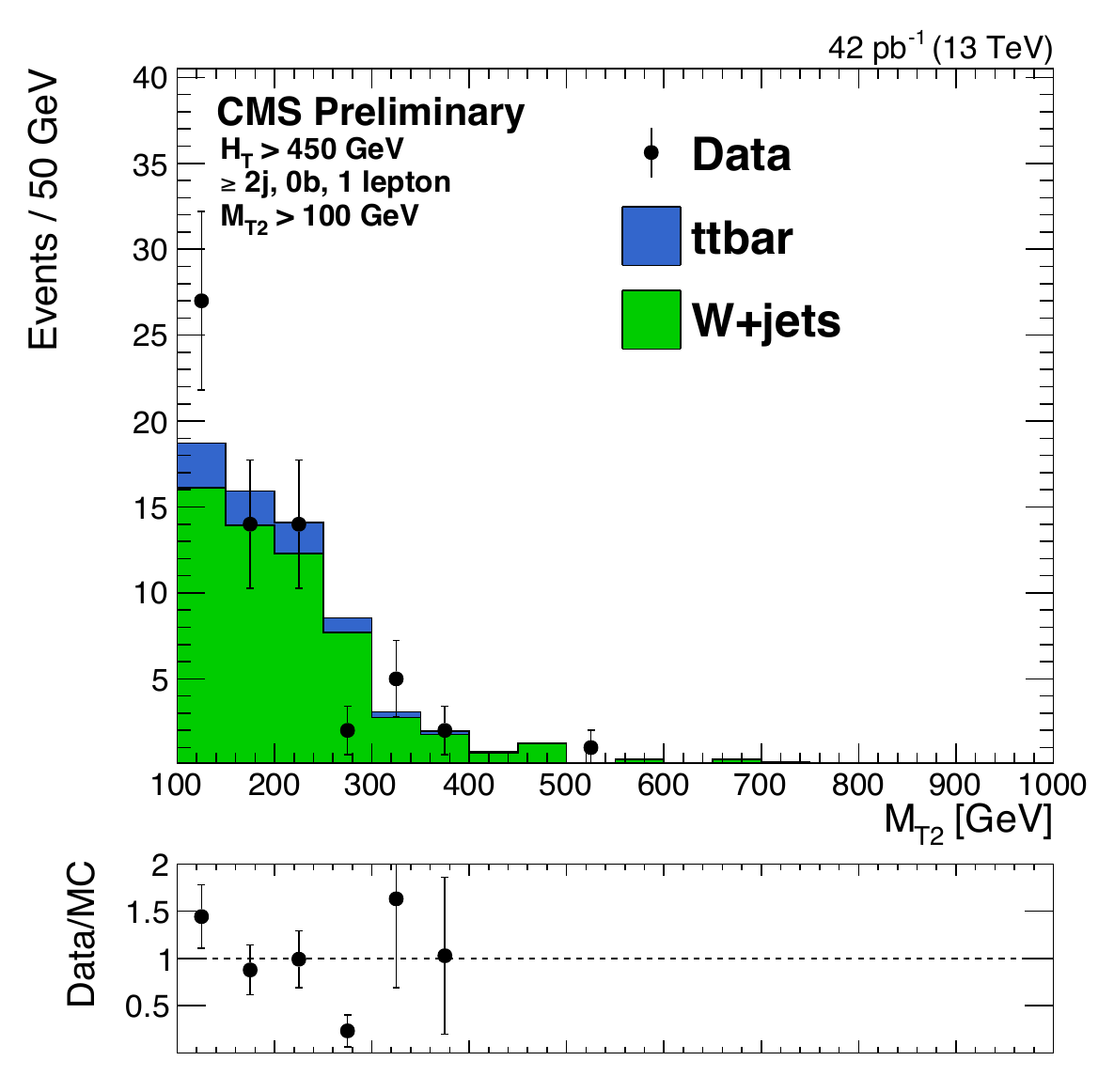}
  \includegraphics[width=0.48\linewidth]{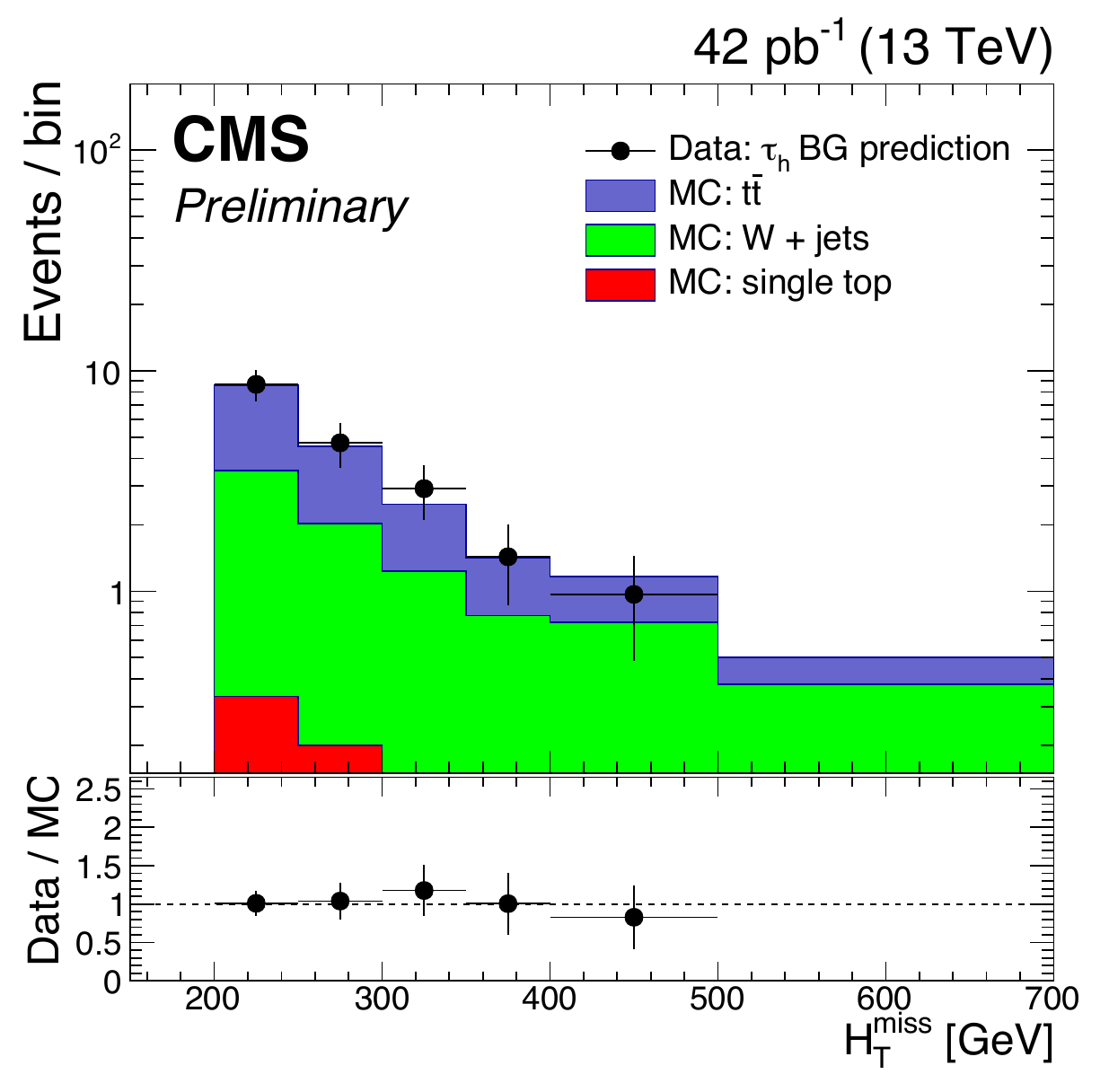}
  \caption{Early 13 TeV data commissioning plots for lost lepton (left) and hadronic tau (right) backgrounds for hadronic SUSY searches.
  Taken from Ref.~\cite{SUS-Commissioning-DPS}.}
  \label{fig:13lostlepton}
\end{figure}

Figure~\ref{fig:13trig} shows the trigger efficiencies as measured in 13 TeV data for triggers based on HT and MET. Such triggers
are utilized for hadronic SUSY searches. Figure~\ref{fig:13lostlepton} shows distributions of sensitive SUSY variables in single 
lepton control samples compared to MC simulation. In the left plot, a sample with no $b$-tagged jets is selected to test the modeling
of the $W$ + jets background, while in the right plot, a single muon sample is selected and visible energy templates are used to 
predicted the hadronic tau background. In both cases, the 13 TeV is is observed to be in good agreement with expectation from MC.

\begin{figure}[!ht]
  \includegraphics[width=0.48\linewidth]{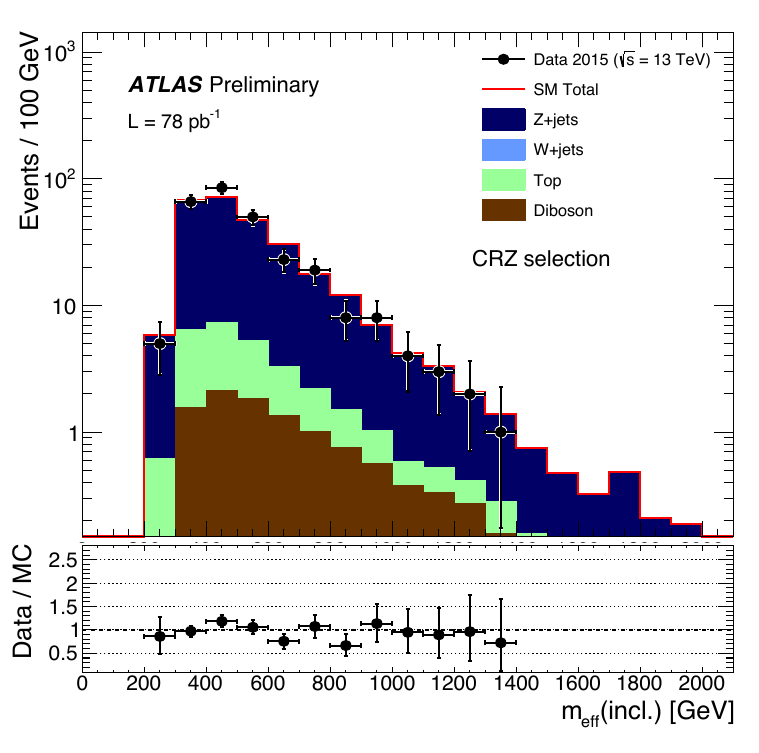}
  \includegraphics[width=0.48\linewidth]{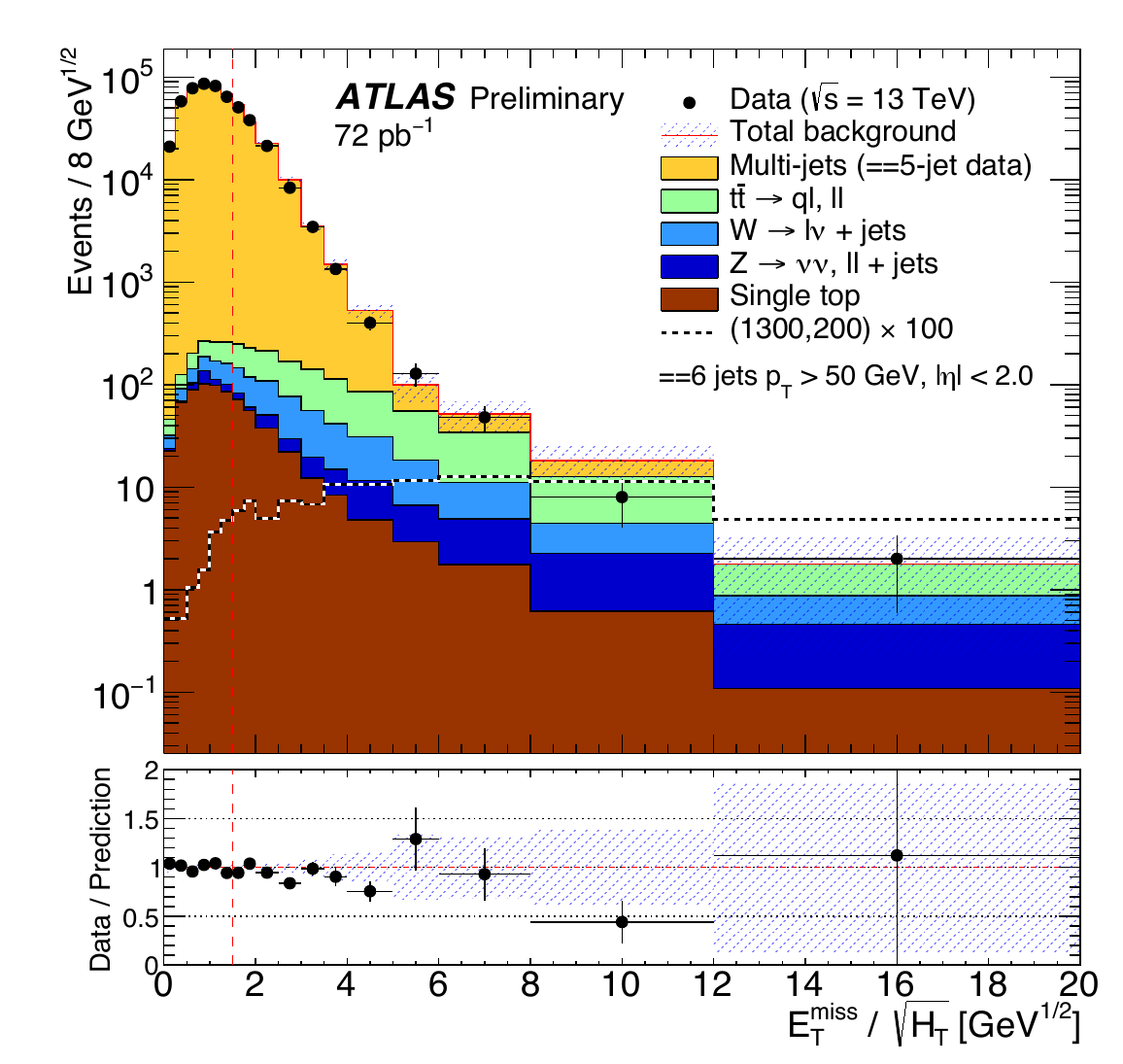}
  \caption{Early 13 TeV data commissioning plots for $Z$ (left) from~\cite{ATLAS-Z}and QCD multijet 
  (right) from ~\cite{ATLAS-QCD} backgrounds for hadronic SUSY searches.}
  \label{fig:13Z-QCD}
\end{figure}

\begin{figure}[!ht]
  \includegraphics[width=0.38\linewidth]{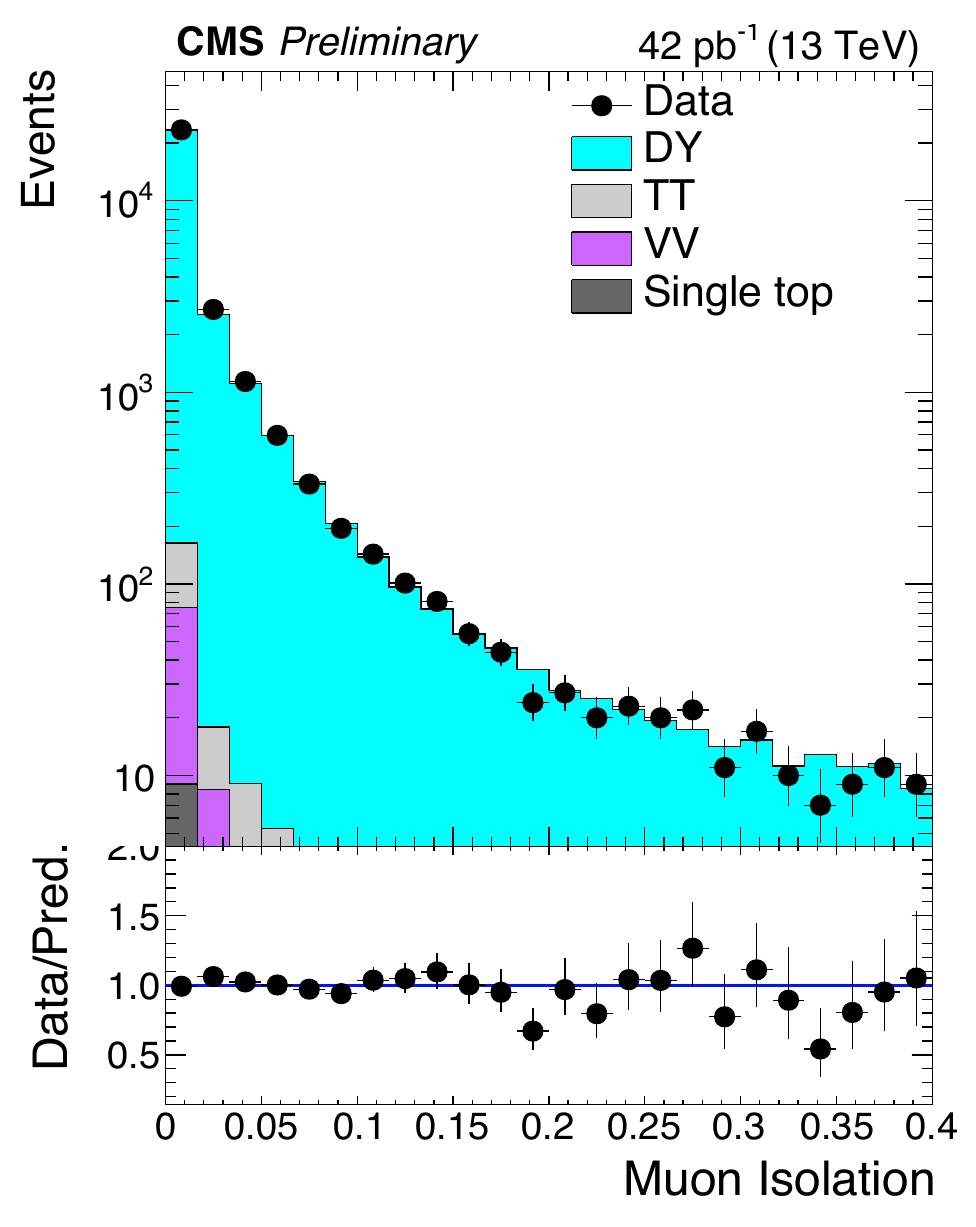}
  \includegraphics[width=0.58\linewidth]{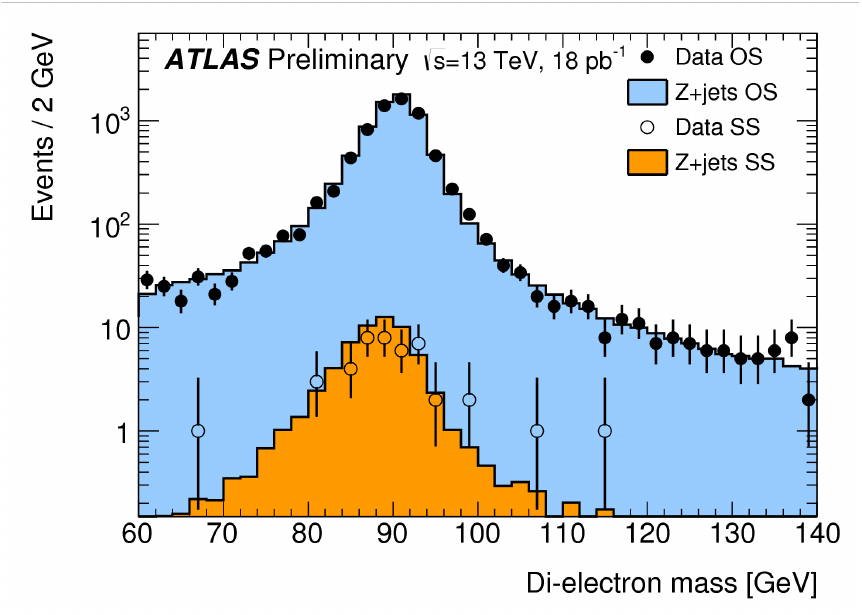}
  \caption{Early 13 TeV data commissioning plots for muon isolation (left) from~\cite{SUS-Commissioning-DPS} and 
  di-electron mass (right) from~\cite{ATLAS-Z} used to 
  predict non-prompt and wrong charge backgrounds, respectively, in same-sign dilepton SUSY searches.}
  \label{fig:13multiLep}
\end{figure}

In Fig.~\ref{fig:13Z-QCD} (left) dilepton control sample events are plotted compared to simulation to test the understanding
of $Z$ + jets background prediction techniques. In Fig.~\ref{fig:13Z-QCD} (right), the distribution of MET/$\sqrt{\textrm{HT}}$ is plotted
comparing 13 TeV data and MC simulation for multijet events, showing good understanding of the QCD multijet background prediction methods.
Figure~\ref{fig:13multiLep} shows 13 TeV data commissioning results for the same-sign dilepton search. In the left plot, the isolation 
distribution for identified muons is plotted, showing good agreement between data and simulation. In the right plot, the di-electron
mass is shown for same-sign and opposite-sign dilepton events, which is a  key ingredient in measuring the rate of wrong charge 
assignment in electrons.

In summary, the 13 TeV data are remarkably well understood only a short period after data taking. Both experiments are well on track
for producing SUSY search results when a sufficient amount of 13 TeV is available. First results are expected based on the full 2015 dataset.

\section{CONCLUSIONS}
Supersymmetry remains among the most popular extensions to the Standard Model. This talk reviews a sample of SUSY results from Run 1 of
the LHC at 8 TeV. The CMS and ATLAS experiments have each produced a large number of SUSY results, with no significant deviations from
Standard Model expectations yet observed. Data taking has begun for Run 2 at 13 TeV. Early commissioning results show that both experiments
are on track to produce exciting new results with this data in the near future. Exciting times are ahead.


\nocite{*}
\bibliographystyle{aipnum-cp}%
\bibliography{sample}%

\section{REFERENCES}

\begin{itemize}

\bibitem{CMS}
CMS Collaboration, JINST {\bf 3}, S08004 (2008).

\bibitem{ATLAS} 
ATLAS Collaboration, JINST {\bf 3}, S08003 (2008).

\bibitem{CMS-SUSY}
CMS Collaboration, ``CMS Supersymmetry Physics Results,'' \url{https://twiki.cern.ch/twiki/bin/view/CMSPublic/PhysicsResultsSUS}.

\bibitem{ATLAS-SUSY}
ATLAS Collaboration, ``ATLAS Supersymmetry searches'' \url{https://twiki.cern.ch/twiki/bin/view/AtlasPublic/SupersymmetryPublicResults}.

\bibitem{Chatrchyan:2013wxa} 
CMS Collaboration, Phys.\ Lett.\ B {\bf 725}, 243 (2013) [arXiv:1305.2390 [hep-ex]].

\bibitem{Aad:2015iea} 
ATLAS Collaboration, JHEP {\bf 1510}, 054 (2015) [arXiv:1507.05525 [hep-ex]].

\bibitem{Chatrchyan:2013fea} 
CMS Collaboration], JHEP {\bf 1401}, 163 (2014)
  [JHEP {\bf 1501}, 014 (2015)] [arXiv:1311.6736, arXiv:1311.6736 [hep-ex]].

\bibitem{Aad:2015hea} 
ATLAS Collaboration, Phys.\ Rev.\ D {\bf 92}, no. 7, 072001 (2015) [arXiv:1507.05493 [hep-ex]].

\bibitem{Aad:2015eda} 
ATLAS Collaboration, arXiv:1509.07152 [hep-ex].

\bibitem{CMS-stop}
CMS Collaboration, CMS-PAS-SUS-13-023, \url{http://cds.cern.ch/record/2044441} (2015).

\bibitem{Aad:2014mfk} 
ATLAS Collaboration, Phys.\ Rev.\ Lett.\  {\bf 114}, no. 14, 142001 (2015) [arXiv:1412.4742 [hep-ex]].

\bibitem{CMS-softL}
CMS Collaboration, CMS-PAS-SUS-14-021, \url{http://cds.cern.ch/record/2010110} (2015).

\bibitem{CMS-VBF}
CMS Collaboration, CMS-PAS-SUS-14-019, \url{http://cds.cern.ch/record/2046988} (2015).

\bibitem{CMS-pMSSM}
CMS Collaboration, CMS-PAS-SUS-15-010, \url{http://cds.cern.ch/record/2063744} (2015).

\bibitem{Aad:2015baa}
ATLAS Collaboration, JHEP {\bf 1510}, 134 (2015) [arXiv:1508.06608 [hep-ex]].

\bibitem{SUS-Commissioning-DPS} CMS Collaboration, ``Commissioning the performance of key observables used in SUSY
searches with the first 13 TeV data,'' CMS-DP-2015-035, \url{http://cds.cern.ch/record/2049757} (2015) .

\bibitem{ATLAS-Z}
ATLAS Collaboration, ATL-PHYS-PUB-2015-028, \url{https://atlas.web.cern.ch/Atlas/GROUPS/PHYSICS/PUBNOTES/ATL-PHYS-PUB-2015-028} (2015).

\bibitem{ATLAS-QCD}
ATLAS Collaboration, ATL-PHYS-PUB-2015-030, \url{https://atlas.web.cern.ch/Atlas/GROUPS/PHYSICS/PUBNOTES/ATL-PHYS-PUB-2015-030} (2015).

\end{itemize}

\end{document}